\documentclass[twocolumn]{aastex63}

\usepackage{natbib}
\usepackage{color}
\usepackage{mathtools}
\usepackage{hyperref} 
\hypersetup{colorlinks=true,citecolor=blue}

\def    \apjl  		{\rm {ApJL}}

\def    \apjl  		{\rm {ApJL}}

\def	\cm		{\,{\rm {cm}}}
\def	\K		{\,{\rm {K}}}
\def	\g		{\,{\rm {g}}}
\def	\mum	{\,{\mu \rm{m}}}

\def \bea {\begin{eqnarray}}
\def \ena {\end{eqnarray}}

\def	\B	{{\rm B}}

\def	\bv	{{\bf v}}

\def	\C	{{\rm C}}

\def	\cm	{\,{\rm cm}}
\def	\d	{{\rm d}}

\def	\D	{{\rm D}}

\def	\erg	{\,{\rm erg}}

\def	\g	{\,{\rm g}}
\def	\gas	{\,{\rm gas}}
\def	\km	{\,{\rm km}}
\def	\G	{{\rm G}}

\def	\H	{{\rm H}}

\def	\N	{{\rm N}}

\def	\s	{\,{\rm s}}

\def	\V	{{\rm V}}

\def	\rad	{\,{\rm rad}}
\def	\yr	    {\,{\rm yr}}

\def    \gas     	{{\rm gas}}

\begin{document}
\shorttitle{Mechanical Torque Disruption}
\shortauthors{Hoang and Lee}
\title{Rotational disruption of dust grains by mechanical torques for high-velocity gas-grain collisions}
\author{Thiem Hoang}
\affil{Korea Astronomy and Space Science Institute, Daejeon 34055, Republic of Korea; \href{mailto:thiemhoang@kasi.re.kr}{thiemhoang@kasi.re.kr}}
\affil{University of Science and Technology, Korea, (UST), 217 Gajeong-ro Yuseong-gu, Daejeon 34113, Republic of Korea}
\author{Hyeseung Lee}
\affil{Korea Astronomy and Space Science Institute, Daejeon 34055, Republic of Korea} 

\begin{abstract}
Dust grains moving at hypersonic velocities of $v_{d}\gtrsim 100\km\s^{-1}$ through an ambient gas are known to be destroyed by nonthermal sputtering. Yet, previous studies of nonthermal sputtering disregarded the fact that dust grains can be spun-up to suprathermal rotation by stochastic mechanical torques from gas-grain collisions. In this paper, we show that such grain suprathermal rotation can disrupt a small grain into small fragments because induced centrifugal stress exceeds the maximum tensile strength of grain material, $S_{\rm max}$. We term this mechanism {\it MEchanical Torque Disruption} (METD). We find that METD is more efficient than nonthermal sputtering in destroying smallest grains ($a<10$ nm) of nonideal structures moving with velocities of $v_{d}<500 \km\s^{-1}$. The ratio of rotational disruption to sputtering time is $\tau_{\rm disr}/\tau_{\rm sp}\sim 0.7(S_{\max}/10^{9}\erg\cm^{-3})(\bar{A}_{\rm sp}/12)(Y_{\rm sp}/0.1)(a/0.01\mum)^{3}(300\km\s^{-1}/v_{d})^{2}$ where $a$ is the radius of spherical grains, and $Y_{\rm sp}$ is sputtering yield. We also consider the high-energy regime and find that the rate of METD is reduced and becomes less efficient than sputtering for $v_{d}>500\km\s^{-1}$ because impinging particles only transfer part of their momentum to the grain. We finally discuss implications of the METD mechanism for the destruction of hypersonic grains accelerated by radiation pressure as well as grains in fast shocks. Our results suggest that the destruction of small grains by METD in fast shocks of supernova remnants may be more efficient than previously predicted by nonthermal sputtering, depending on grain internal structures. 
 
\end{abstract}

\section{Introduction}
The motion of dust grains at high velocities above $\sim 100 \km\s^{-1}$ through the ambient gas (hereafter hypersonic motion) is common in the universe. Various physical processes can accelerate dust grains to hypersonic velocities, including radiation pressure induced by strong radiation sources (e.g., late-type stars, supernovae, and active galactic nuclei; \citealt{1976ApJ...205..144G}; \citealt{1993ApJ...410..701N}) and shock waves. Moreover, charged grains can be accelerated to high velocities in interstellar shocks of supernova remnants via betatron and Fermi acceleration mechanisms (\citealt{1980MNRAS.193..723E}; \citealt{1997ApJ...487..197E}). Magnetohydrodynamic turbulence is found to accelerate charged dust grains to high velocities (\citealt{Yan:2004ko}; \citealt{Hoang:2012cx}). In particular, newly formed dust grains in the supernova ejecta move hypersonically through the ambient gas before injected into the diffuse interstellar medium.

Hypersonic grain motion is thought to play an important role for a wide range of astrophysical phenomena such as galactic winds (e.g., \citealt{Ishibashi:2015bu}), dust transport from the galaxy to the circumgalactic and intergalactic medium (\citealt{1991ApJ...381..137F}; \citealt{2001ApJ...556L..11A}; \citealt{2001ApJ...561..521A}; \citealt{2005MNRAS.358..379B}). Therefore, the critical question is whether dust grains can survive during the transport from the galaxy into the intergalactic medium (IGM). The survival of dust in the supernova ejecta is crucially important for quantifying the dust budget in the universe. 

Nonthermal sputtering is believed to be a dominant mechanism for destruction of hypersonic grains (\citealt{1979ApJ...231..438D}). Thus, understanding the physics of sputtering is critically important for quantitative understanding of the formation and destruction of cosmic dust (see, e.g., \citealt{2006ApJ...648..435N}). The underlying physics of sputtering is that an impinging ion/atom can eject target atoms from the grain surface via nuclear-nuclear or electronic interactions (\citealt{1981spb1.book....9S}). 

In fast shocks of velocities $v_{\rm sh}\gtrsim 100 \km\s^{-1}$, nonthermal sputtering is usually referred to explain the destruction of dust grains (\citealt{1979ApJ...231..438D}; \citealt{1994ApJ...433..797J}; \citealt{2010ApJ...715.1575S}). Observations reveal dust destruction in SNRs (\citealt{2015ApJ...799...50L}). Nevertheless, observations show that the fraction of dust destroyed in fast shocks is higher than predicted by theoretical predictions based on sputtering (\citealt{2006ApJ...652L..33W}; \citealt{2010ApJ...712.1092S}; \citealt{2019arXiv190706213Z}). This motivates us to look for physical effect ignored in the current theory of sputtering.

Indeed, previous studies of nonthermal sputtering disregarded the fact that the grain can be spun-up to suprathermal rotation by stochastic gas-grain collisions as pointed out by \cite{1952MNRAS.112..215G} and numerically demonstrated through Monte Carlo simulations by \cite{1971ApJ...167...31P}. Recently, \cite{2019ApJ...877...36H} studied the effect of gas bombardment for nanoparticles drifting in steady-state shocks of low velocities of $v_{\rm sh}<50 \km\s^{-1}$. The authors found that the smallest nanoparticles (size below 2 nm) can be spun-up to suprathermal rotation by stochastic mechanical torques and destroyed by centrifugal stress. This mechanism can be termed MEchanical Torque Disruption (METD). {\it The question is whether METD is still efficient for gas-grain collisions at higher velocities.}

The key difference between low and high-velocity gas-grain collisions is that, at high velocities (i.e., $v\gtrsim 50 \km\s^{-1}$), incident particles may pass through instead of stick to the grain when the particle penetration length exceeds the grain diameter. As a result, they do not transfer their entire momentum to the grain upon collisions  (\citealt{2017ApJ...847...77H}), and the efficiency of METD is reduced. The goal of this paper is to quantify the efficiency of METD for both low-energy (i.e., stick) and high-energy (passage) regimes and compare with nonthermal sputtering.

The structure of our paper is as follows. In Section \ref{sec:disr}, we study the spin-up and rotational disruption of grains by stochastic mechanical torques upon gas-grain collisions, and we compare METD with nonthermal sputtering in Section \ref{sec:compare}. Section \ref{sec:discuss} is devoted to discussing the implications of our study for dust destruction in fast shocks and grains accelerated by radiation pressure and its transport into the IGM. The summary of our main findings is presented in Section \ref{sec:summary}.

\section{Rotational disruption by stochastic torques from gas-grain collisions}\label{sec:disr}
\subsection{General consideration}
We consider a spherical grain of radius $a$ moving through the ambient gas of atomic hydrogen of temperature $T_{\rm gas}$ and number density $n_{\rm H}$. Let define isothermal Mach sonic number $s_{d}=v_{d}/v_{T}$ where $v_{T}=(2kT_{\rm gas}/m_{\rm H})^{1/2}\simeq 1.2 T_{2}^{1/2}~\km\s^{-1}$ with $T_{2}=T_{\rm gas}/(100\K)$ is the thermal gas velocity. For the supersonic regime of $s_{d}\gg 1$ considered in this paper, the effect of thermal (Brownian) collisions is subdominant and can be ignored.

We first consider the {\it low-energy regime} where incident species collide and stick to the grain surface, such that they transfer their entire momentum to the grain. We then consider {\it high energy regime} where the penetration length of particles is larger than the grain diameter, such that they can pass through the grain, transferring part of their momentum to the grain \citep{2017ApJ...847...77H}. 

\subsection{Low-energy regime}
\subsubsection{Spin-up by stochastic torques from gas-grain collisions}

Let us estimate the rotational excitation of grains due to sticking collisions of gas species. Each atom colliding with the grain surface at radius ${\bf r}$ transfers its entire momentum $m_{\H}v$ to the grain, inducing an {\it impulsive torque} of $\delta {\bf J}= {\bf r}\times m_{\H}\bv$ (see e.g., \citealt{1952MNRAS.112..215G}). The increase of $(\delta J)^{2}$ from each impact is given by
\bea
(\delta J)^{2} = (a\cos\theta m_{\H}v_{d})^{2}= m_{\H}^{2}v_{d}^{2}a^{2}\cos^{2}\theta,\label{eq:dJsqr}
\ena
where $\theta$ is the polar angle of the radius vector ${\bf r}$, and the projectiles are impinging along the horizontal plane.

By averaging the above equation over the grain surface, one has $\langle \cos^{2}\theta\rangle= 1/2$. Thus, Equation (\ref{eq:dJsqr}) becomes
\bea
\langle(\delta J)^{2}\rangle = \frac{1}{2}m_{\H}^{2}v_{d}^{2}a^{2},\label{eq:dJ_sqr}
\ena
which yields the rotational angular velocity acquired by a single collision: 
\bea
\delta \omega =\frac{\langle(\delta J)^{2}\rangle^{1/2}}{I} \simeq 1.6\times 10^{7}a_{-6}^{-4}v_{3}\rad\s^{-1},\label{eq:delta_omega}
\ena
where the inertia moment $I=8\pi \rho a^{5}/15$ with $\rho$ the grain mass density, $a_{-6}=a/(10^{-6}\cm)$, and $v_{3}=v_{d}/(10^{3}\km\s^{-1})$.

Using the random walk theory for stochastic collisions, one can derive the total increase of squared angular momentum per unit of time as follows:
\bea
\frac{\langle(\Delta J)^{2}\rangle}{\Delta t} = R_{\rm coll}(\delta J)^{2} = \frac{n_{\H}v_{d} \pi a^{2}m_{\H}^{2}v_{d}^{2}a^{2}}{2},\label{eq:dJ2}
\ena
where the collision rate $R_{\rm coll}= n_{\H}v_{d} \pi a^{2}$ has been used.

After traversing a time interval $\Delta t$, the total average increase of the squared angular momentum is equal to
\bea
\langle(\Delta J)^{2}\rangle = \frac{n_{\rm H}m_{\H}^{2}v_{d}^{3}\pi a^{4}}{2}\Delta t.\label{eq:deltaJ}
\ena

The {\it rms} angular velocity of grains can now be calculated using the total angular momentum $\Delta J$ from Equation (\ref{eq:deltaJ}):
\bea
\omega_{\rm rms}^{2}=\langle \omega^{2}\rangle &=& \frac{\langle(\Delta J)^{2}\rangle}{I^{2}}=\frac{n_{\rm H}m_{\H}^{2}v^{3}\pi a^{4}}{2I^{2}}\Delta t.\label{eq:omegasqr}
\ena

A rotating grain experiences rotational damping due to sticking collision with gas atoms. Note that sticking collisions do not damp grain rotation due to averaging effect, but subsequent thermal evaporation of atoms that carry away part of the grain angular momentum results in grain rotational damping (see e.g., \citealt{1998ApJ...508..157D}). Consider a grain rotating along the $z$-axis with angular velocity $\omega_{z}$. The angular momentum carried away by an H atom from the grain surface is given by
\bea
\delta J_{z} = I_{m}\omega_{z} = m_{\H}r^{2}\omega_{z}= m_{\H}a^{2}\sin^{2}\theta\omega_{z},
\ena
where $r$ is the distance from the atom to the spinning axis $z$, $I_{m}=m_{\H}r^{2}$ is the inertial moment of the hydrogen atom of mass $m_{\H}$, $\theta$ is the angle between the z-axis and the radius vector, and $r=a\sin\theta$ is the projected distance to the center. Assuming the isotropic distribution of $\theta$ for atoms leaving the grain, one can replace $\sin^{2}\theta = <\sin^{2}\theta> = 2/3$, which give rise to

\bea
\langle \delta J_{z}\rangle = \frac{2}{3}m_{\H}a^{2}\omega_{z}.
\ena

Using the collision rate of atomic gas, $R_{\rm coll}$, one can derive the mean decrease of grain angular momentum per unit of time is
\bea
\bigg\langle\frac{\Delta J_{z}}{\Delta t}\bigg\rangle_{\H}&=& -R_{\rm coll}\langle \delta J_{z}\rangle\nonumber\\
&=&-\frac{2}{3}n_{\H}m_{\H}\pi a^{4}\omega_{z}\langle v\rangle=-\frac{I\omega_{z}}{\tau_{\rm H}}.
\ena

For the drift velocity with $v\gg v_{T}$, one has $\langle v\rangle =v_{d}$. Therefore, the rotational damping time is
\bea
\tau_{\rm H}&=&\frac{3I}{2n_{\rm H}m_{\rm H}\pi a^{4}v_{d}}=\frac{4\rho a}{5n_{\rm H}m_{\rm H}v_{d}}\simeq 572\left(\frac{\hat{\rho} a_{-6}}{v_{2}n_{1}}\right)~\rm yr,~~~~~\label{eq:tauH}
\ena
where $n_{1}=n/(10\cm^{-3}), \hat{\rho}=\rho/(3\g\cm^{-3}), v_{2}=v_{d}/(100\km\s^{-1})$.

Rapidly spinning dust grains emit strong electric dipole radiation (\citealt{1998ApJ...508..157D}), which also damps the grain rotation on a timescale of
\bea
\tau_{\rm ed}=\frac{3I^{2}c^{3}}{\mu^{2}kT_{\gas}}\simeq  2.25\times 10^{8} \left(\frac{a_{-6}^{7}}{3.8\hat{\beta}}\right)\left(\frac{100\K}{ T_{\gas}}\right)\rm yr,\label{eq:taued}
\ena
where $\mu$ is the grain dipole moment and $\hat{\beta}=\beta/(0.4\D)$ with $\beta$ being the dipole moment per structure due to polar bonds present in the dust grain (\citealt{1998ApJ...508..157D}; \citealt{Hoang:2010jy}; \citealt{2016ApJ...824...18H}). 

Comparing $\tau_{\rm ed}$ with $\tau_{\H}$, one can see that, for small grains of $a>1$ nm, the electric dipole damping time is longer than the gas damping time. 

Due to the rotational damping, the grain looses angular momentum on a timescale of $\tau_{\rm H}$. Therefore, Equation (\ref{eq:omegasqr}) yields
\bea
\omega_{\rm rms}^{2}\equiv \langle \omega^{2}\rangle &=& \frac{n_{\rm H}m_{\H}^{2}v^{3}\pi a^{4}}{2I^{2}}\tau_{\rm H},\label{eq:omegasqr2}
\ena
which can be rewritten as
\bea
\frac{\omega_{\rm rms}^{2}}{\omega_{T}^{2}}=\frac{s_{d}^{2}}{2},
\ena
where the thermal angular velocity
\bea
\omega_{T}&=&\left(\frac{3kT_{\rm gas}}{I}\right)^{1/2}\nonumber\\
&\simeq& 9\times 10^{7}a_{-6}^{-5/2}T_{2}^{1/2}\hat{\rho}^{-1/2}~ \rad\s^{-1}.
\ena

\subsubsection{Mechanical Torque Disruption Mechanism}
The basic idea of rotational disruption by stochastic mechanical torques (i.e., METD mechanism), is as follows. A spherical dust grain rotating at velocity $\omega$ develops a centrifugal stress due to centrifugal force, which is maximum along the plane through the grain center of $S=\rho a^{2} \omega^{2}/4$ (\citealt{Hoang:2019da}). When the rotation rate increases to a critical limit such that the tensile stress induced by centrifugal force exceeds the maximum tensile stress, the so-called tensile strength of the material, the grain is disrupted instantaneously. The critical angular velocity for the disruption is given by
\bea
\omega_{\rm cri}=\frac{2}{a}\left(\frac{S_{\rm max}}{\rho} \right)^{1/2}
\simeq 3.65\times 10^{10}\left(\frac{S_{\rm max,9}^{1/2}}{a_{-6}\hat{\rho}^{1/2}}\right)\rm rad\s^{-1},~~~~\label{eq:omega_cri}
\ena
where $S_{\rm max}$ is the tensile strength of dust material and $S_{\rm max,9}=S_{\rm max}/(10^{9}\erg\cm^{-3})$ is the tensile strength in units of $10^{9}\erg\cm^{-3}$. An alternative unit of the tensile strength is ${\rm dyne/cm^{2}}$, but in this paper we use the unit of $\erg\cm^{-3}$ for $S_{\rm max}$. 

The tangential velocity required for the disruption is 
\bea
v_{\rm cri}\sim \omega_{\rm cri} a\sim 0.36\left(\frac{S_{\rm max,9}^{1/2}}{\hat{\rho}^{1/2}}\right) \rm km\s^{-1},
\ena
which is much smaller than the gas thermal velocity of $v_{T}\sim 1.2 T_{2}^{1/2}~\km\s^{-1}$.

The exact value of $S_{\max}$ depends on grain composition and structure. Compact grains are expected to have higher $S_{\max}$ than porous/composite grains. Ideal material without impurity, such as diamond, can have $S_{\max}\ge 10^{11}\erg\cm^{-3}$ (see \citealt{Hoang:2019da} for more details). In the following, grains with $S_{\max} \gtrsim 10^{9}\erg\cm^{-3}$ are referred to as strong materials, and those with $S_{\max}< 10^{9}\erg\cm^{-3}$ are called weak materials. 

\subsubsection{Disruption time and disruption size}
The time required to spin-up a grain of size $a$ to $\omega_{\rm cri}$, so-called rotational disruption time, is evaluated as follows:
\bea
\tau_{\rm disr}&=&
\frac{J_{\rm cri}^{2}}{(\Delta J)^{2}/(\Delta t)}=\frac{2(I\omega_{\rm cri})^{2}}{n_{\H}m_{\H}^{2}v_{d}^{3}\pi a^{4}}= \frac{512\pi a^{4}\rho S_{\max}}{225n_{\H}m_{\H}^{2}v_{d}^{3}}\nonumber\\
&\simeq & 2.4\times 10^{4}\left(\frac{a_{-6}^{4}}{v_{2}^{3}}\right)\left(\frac{S_{\rm max,9}}{n_{1}\hat{\rho}}\right){\rm yr}~~~.\label{eq:tdisr}
\ena

The above equation implies that nanoparticles of $a\sim 1$ nm moving at $v_{d}\sim 100 \km\s^{-1}$ are disrupted in $t_{\rm disr}\sim 2$ yr, while the grain rotation is damped in $\tau_{\H}\sim 50$ yr by gas collisions or in $\tau_{\rm ed}\sim 20$ yr by electric dipole emission.

We note that METD only occurs when the required time is shorter than the rotational damping time. Let $a_{\rm disr}$ be the grain disruption size as determined by $\tau_{\rm disr}=\tau_{\H}$. Thus, comparing Equations (\ref{eq:tdisr}) and (\ref{eq:tauH}), one obtains:
\bea
a_{\rm disr}=\left(\frac{25m_{\H}v_{d}^{2}}{128\pi S_{\rm max}}\right)^{1/3}\simeq 5.5S_{\max,9}^{-1/3}\left(\frac{v_{d}}{300\rm km\s^{-1}}\right)^{2/3} ~~\rm nm,~~~\label{eq:adisr}
\ena
which implies that very small grains ($a<5$ nm) moving at $v_{d}\sim 300\rm km\s^{-1}$ are disrupted by centrifugal stress, assuming strong grains of $S_{\max}\sim 10^{9}\erg\cm^{-3}$. The rotation of larger grains (i.e., $a>a_{\rm disr}$) is damped by gas collisions before reaching the critical threshold.

For a given grain size, the critical speed required for rotational disruption is given by the condition of $\tau_{\rm disr}\lesssim \tau_{\H}$, which yields
\bea
v_{d}\gtrsim \left(\frac{128\pi a^{3}S_{\rm max}}{45m_{\rm H}}\right)^{1/2}\simeq 733a_{-6}^{3/2}S_{\max,9}^{1/2} \km\s^{-1}.~~~\label{eq:vdisr}
\ena
or the dimensionless parameter:
\bea
s_{d}\gtrsim \left(\frac{64\pi a^{3}S_{\rm max}}{45kT_{\rm gas}}\right)^{1/2}\simeq 565a_{-6}^{3/2}T_{2}^{-1/2}S_{\max,9}^{1/2}.\label{eq:sdisr}
\ena

The above equations indicate that the velocity required for METD decreases rapidly with decreasing grain size and with tensile strength. Smallest nanoparticles of sizes $a\sim 1$ nm only require $v_{d}\sim 23\km\s^{-1}$ while small grains of $a\sim 0.01\mum$ require much higher velocities for rotational disruption, assuming $S_{\rm max}\lesssim 10^{9}\erg\cm^{-3}$.

The rotation of nanoparticles experiences damping and excitation by various interaction processes, including ion collisions, plasma drag, and infrared emission (see \citealt{1998ApJ...508..157D}; \citealt{Hoang:2010jy}). A detailed analysis of the different damping processes for grains in magnetized shocks is presented in \cite{2019ApJ...877...36H} and \cite{2019ApJ...886...44T}.

\subsubsection{Slowing-down time by gas drag force}
For hypersonic grains, the main gas drag arises from direct collisions with gas atoms, and the Coulomb drag force is subdominant (\citealt{1979ApJ...231...77D}). Assuming the sticky collisions of atoms followed by their thermal evaporation, the decrease in the grain momentum is equal to the momentum transferred to the grain in the opposite direction:
\bea
F_{\rm drag}\equiv \frac{dP}{dt} = m_{\H}v_{d}\times n_{\rm H}v_{d}\pi a^{2}.
\ena

The gas drag time is given by 
\bea
\tau_{\rm drag}&=& \frac{m_{gr}v_{d}}{dP/dt}=\frac{4\pi \rho a^{3}v_{d}}{3\pi a^{2}n_{\rm H}v_{d}^{2} m_{\rm H}}\nonumber\\
 &=&\frac{4\rho a}{3n_{\rm H}m_{\rm H}v_{d}}\simeq 763\left(\frac{\hat{\rho}a_{-6}}{n_{1}v_{2}}\right)~\yr.\label{eq:tdrag}
\ena

Comparing Equations (\ref{eq:tdrag}) with (\ref{eq:tdisr}) one can see that the disruption occurs much faster than the drag time for $v>100\km\s^{-1}$ and small grains of $a<0.01\mum$.

\subsection{High-energy regime}
The penetration depth of impinging protons is approximately equal to (\citealt{1979ApJ...231...77D}):
\bea
R_{\rm H}(E)\simeq \left(\frac{0.01}{\hat{\rho}}\right)\left(\frac{E}{1~\rm keV}\right) \mum\simeq 0.008 \left(\frac{v_{2}^{2}}{\hat{\rho}}\right)\mum,~~~\label{eq:RH}
\ena
which reveals that for high-velocity collisions, impinging particles can pass through the grain because $R_{\rm H}> 2a$. As a result, they only transfer part of their momentum to the grain. We will first find the fraction of ion momentum transferred to the grain and quantify the efficiency of METD.

For interstellar grains with $a<1\mum$ and energetic ions, we have $\Delta E \ll p^{2}/2m$, \cite{2017ApJ...847...77H} derived
\begin{eqnarray}
\delta p =  \frac{2mp\delta E}{2p^{2} - m\delta E} \approx  p\left(\frac{\delta E}{2E}\right)=p f_{p}(E,a),\label{eq:dp}
\end{eqnarray}
where $\Delta E$ is the energy loss passing the grain, and $f_{p}(E,a)=\delta E/(2E)$ is the fraction of the ion energy transferred to the grain which is a function of $E$ and $a$.

Let $dE/dx=nS(E)$ where $S(E)$ be the stopping cross-section of the impinging ion of kinetic energy $E$ in the dust grain of atomic density $n$ \citep{1981spb1.book....9S}. The energy loss of the ion due to the passage of the grain is given by
\bea
\delta E= \frac{4a}{3}nS(E),~\label{eq:deltaE}
\ena
where the grain is approximated as slab of thickness $4a/3$. Thus,
\bea
f_{p}=\frac{2a}{3E}nS(E),\label{eq:fE}
\ena
and $f_{p}(E,a)=1$ for sticking collisions.

{\bf Following Equation (\ref{eq:dJ_sqr})}, the impulsive angular momentum from a collision is then given by
\bea
(\delta J)^{2}=\frac{a^{2}}{2}(\delta p)^{2}=\frac{a^{2}p^{2}}{2}f_{p}(E,a)^{2}.
\ena
which yields the average value
\bea
\langle (\delta J)^{2}\rangle = \frac{a^{2}p^{2}}{2}f_{p}(E,a)^{2}.\label{eq:dJ2_hi}
\ena

Following the similar procedure as in Section \ref{sec:disr}, one obtain 
\bea
\langle\frac{(\Delta J)^{2}}{\Delta t}\rangle = \langle\frac{(\Delta J)^{2}}{\Delta t}\rangle_{S}f_{p}^{2},\label{eq:DJ2_hi}
\ena
where $S$ denotes sticking collisions considered in the previous subsection, and $\langle\frac{(\Delta J)^{2}}{\Delta t}\rangle_{S}$ is given by Equation (\ref{eq:dJ2}).

The increase of the grain angular velocity is given by 
\bea
\omega_{\rm rms}^{2}=\frac{(\Delta J)^{2}/\Delta t}{I^{2}}\times t=\left(\frac{n_{\H}m_{\H}v^{3}\pi a^{4}}{2I^{2}}\right)f_{p}^{2}\times t.~~~
\ena

If the incident ion passes through the grain, the grain rotational damping by gas collisions is not important, and the damping by electric dipole emission takes over. Since $\tau_{\rm ed}$ is rather long for nanoparticles of $a>1$ nm (see \citealt{Hoang:2010jy}), the grain angular velocity continues to increase to the critical limit, i.e., at $\omega_{\rm disr}$, i.e., the disruption occurs, in disruption time equal to
\bea
\tau_{\rm disr}=\left(\frac{2I\omega_{\rm disr}^{2}}{n_{\H}m_{\H}v^{3}\pi a^{4}}\right)\frac{1}{f_{p}^{2}}=\tau_{\rm disr,S}\left(\frac{1}{f_{p}^{2}}\right),\label{eq:tdisr_high}
\ena
where $\tau_{\rm disr,S}$ is the disruption time for sticky collisions given by Equation (\ref{eq:tdisr}). The rate of rotational disruption is decreased rapidly with $E$ when $f_{p}<1$. In the above analysis, we assumed a slab model to calculate the fraction of the momentum transfer. As shown in Appendix \ref{apdx:slab}, this assumption induces a negligible difference to the more detailed treatment.

For grain velocity below the Bohr velocity of $v_{0}=e^{2}/\hbar\approx c/137\approx 2189\km\s^{-1}$, nuclear interactions dominate, and the stopping cross-section in units of erg cm$^{2}$ is given by (\citealt{1981spb1.book....9S})
\bea
S_{n}(E) = 4.2\pi a Z_{1}Z_{2}e^{2}\frac{M_{1}}{(M_{1}+M_{2})}s_{n}(\epsilon_{12}),
\ena
where $M_{i}$ and $Z_{i}$ are the atomic masses and numbers charge of the projectile ($i=1$) and target ($i=2$) atom, and $a$ is the screen length for the nuclei-nuclei interaction potential given by
\bea
a\simeq 0.885a_{0}(Z_{1}^{2/3}+Z_{2}^{2/3})^{-1/2},~~a_{0}=0.529\AA,
\ena
and 
\bea
\epsilon_{12}=\left(\frac{M_{2}E}{M_{1}+M_{2}}\right)\left(\frac{a}{Z_{1}Z_{2}e^{2}}\right).
\ena

We adopt the approximate function of $s_{n}$ as in \cite{1994ApJ...431..321T}:
\bea
s_{n}=\frac{3.441\sqrt{\epsilon_{12}}\ln(\epsilon_{12}+2.718)}{1+6.35\sqrt{\epsilon_{12}}+\epsilon_{12}(-1.708+6.882\sqrt{\epsilon_{12}})}.
\ena

For grain velocities above $v_{0}$, electronic interactions dominate, and the stopping power can be approximated as
\begin{eqnarray}
nS_{e}(E) \approx \frac{2nS_{m}(E/E_{m})^{\eta}}{1+(E/E_{m})},\label{eq:dEdx_aprx}
\end{eqnarray}
where $\eta$ is the slope, $E_{m}=100$ keV and $S_{m}$ is the stopping power at $E=E_{m}$. For graphite, we find that $\eta=0.2$ and $nS_{m}= 1.8\times 10^{6}$ keV/cm. For quartz material, $\eta=0.25$ and $nS_{m}=1.3\times 10^{6}$ keV/cm.


Drag force in the high-velocity regime is given by (see also \citealt{2017ApJ...847...77H})
\bea
F_{\rm drag}=R_{\rm coll}\delta p=n_{\H}\pi a^{2}m_{\H}v_{d}^{2}\left(\frac{1}{2 f_{p}}\right).
\ena
The drag time is equal to
\bea
\tau_{\rm drag}=\frac{m_{gr}v_{d}}{F_{\rm drag}}=\tau_{\rm drag,S}\left(\frac{1}{2f_{p}}\right),\label{eq:tdrag_hi}
\ena
where $\tau_{\rm drag,S}$ is given by Equation (\ref{eq:tdrag}).

\subsection{Nonthermal sputtering}
To prepare for comparison, we discuss here nonthermal sputtering which is previously thought to be an efficient mechanism to destroy small grains upon gas-grain collisions (e.g., \citealt{1979ApJ...231..438D}; \citealt{1994ApJ...433..797J}). 

Let $Y_{\rm sp}$ be the average sputtering yield per impinging atom (i.e., H and He) with velocity $v_{d}$. Let $dN_{\rm sp}$ be the number of target atoms sputtered by the bombardment per second, which is given by 
\bea
\frac{dN_{\rm sp}}{dt}=n_{\rm H}v_{d}\pi a^{2}Y_{\rm sp}.
\ena

The rate of grain mass loss due to nonthermal sputtering is given by (see e.g., \citealt{2015ApJ...806..255H})
\begin{eqnarray}
\frac{4\pi \rho a^{2}da}{dt} = m_{\rm sp}\frac{dN_{\rm sp}}{dt}=\bar{A}_{\rm sp}m_{\rm H}n_{\rm H}v_{d}\pi a^{2}Y_{\rm sp},
\end{eqnarray}
which yields
\bea
\frac{da}{dt}&=&\frac{n_{\H}m_{\H}v_{d}Y_{\rm sp}\bar{A}_{\rm sp}}{4\rho},\nonumber\\
&\simeq&5.2\times 10^{-6}~\left(\frac{\bar{A}_{\rm sp}}{12}\right) \left(\frac{n_{1}v_{2}}{\hat{\rho}}\right)\left(\frac{Y_{\rm sp}}{0.1}\right)~\frac{\mu m}{\rm yr},
\ena
where $m_{\rm sp}=\bar{A}_{\rm sp}m_{\H}$ with $\bar{A}_{\rm sp}$ the average atomic mass number of sputtered atoms. Above, we consider the projectiles along the normal vector only.

The sputtering yield, $Y_{\rm sp}$, depends on projectile energy and properties of grain material. Following \cite{1994ApJ...431..321T}, the sputtering yield is given by
\bea
Y_{\rm sp}(E)&=&4.2\times 10^{14}\frac{\alpha S_{n}(E)}{U_{0}}\left(\frac{R_{p}}{R}\right)\nonumber\\
&&\times\left[1-(E_{\rm th}/E)^{2/3}\right]\left(1-E_{\rm th}/E\right)^{2},
\ena
where $U_{0}$ is the binding energy of dust atoms, $\alpha\simeq 0.3(M_{2}/M_{1})^{2/3}$  for $0.5<M_{2}/M_{1}<10$, $\alpha\approx 0.1$ for $M_{2}/M_{1}<0.5$, and $E_{\rm th}$ is the threshold energy for sputtering given by 
\bea
E_{\rm th}&=&\frac{U_{0}}{g(1-g)} {~\rm for ~} M_{1}/M_{2}\le 0.3,\\
E_{\rm th}&=&8U_{0}\left(\frac{M_{1}}{M_{2}}\right)^{1/3} {~\rm for ~} M_{1}/M_{2}> 0.3,
\ena
and $g=4M_{1}M_{2}/(M_{1}+M_{2})^{2}$ is the maximum energy transfer of a head-on elastic collision. The factor $R_{p}/R$ is the ratio of the mean projected range to the mean penetrated path length, as given by \citep{1984NIMPB...2..587B}
\bea
\frac{R_{p}}{R}=\left(K\frac{M_{2}}{M_{1}}+1 \right)^{-1}
\ena
where $K$ is a free parameter, and $K=0.1$ and $0.65$ for silicate and graphite grains, respectively (see \citealt{1994ApJ...431..321T}).

The characteristic timescale of grain destruction by nonthermal sputtering for a grain of size $a$ is defined by 
\bea
\tau_{\rm sp}&=&\frac{a}{da/dt}=\frac{4\rho a}{n_{\H}m_{\H}v_{d}Y_{\rm sp}\bar{A}_{\rm sp}}\nonumber\\&\simeq& 1.9\times 10^{3}\hat{\rho}\left(\frac{12}{\bar{A}_{\rm sp}}\right)\left(\frac{a_{-6}}{n_{1}v_{2}}\right) \left(\frac{0.1}{Y_{\rm sp}}\right) \rm yr.\label{eq:tau_sp}
\ena

\section{Numerical results}\label{sec:compare}
For numerical results shown in this section, we assume graphite grains ($\rho=2.2\g\cm^{-3}$, $U_{0}=4$ eV) moving through purely hydrogen gas with gas density $n_{\H}=10\cm^{-3}$ and $T_{\gas}=100\K$. The present theory can be applied to arbitrary grain materials and astrophysical environments.

\subsection{Grain disruption time}
\begin{figure}
\includegraphics[width=0.5\textwidth]{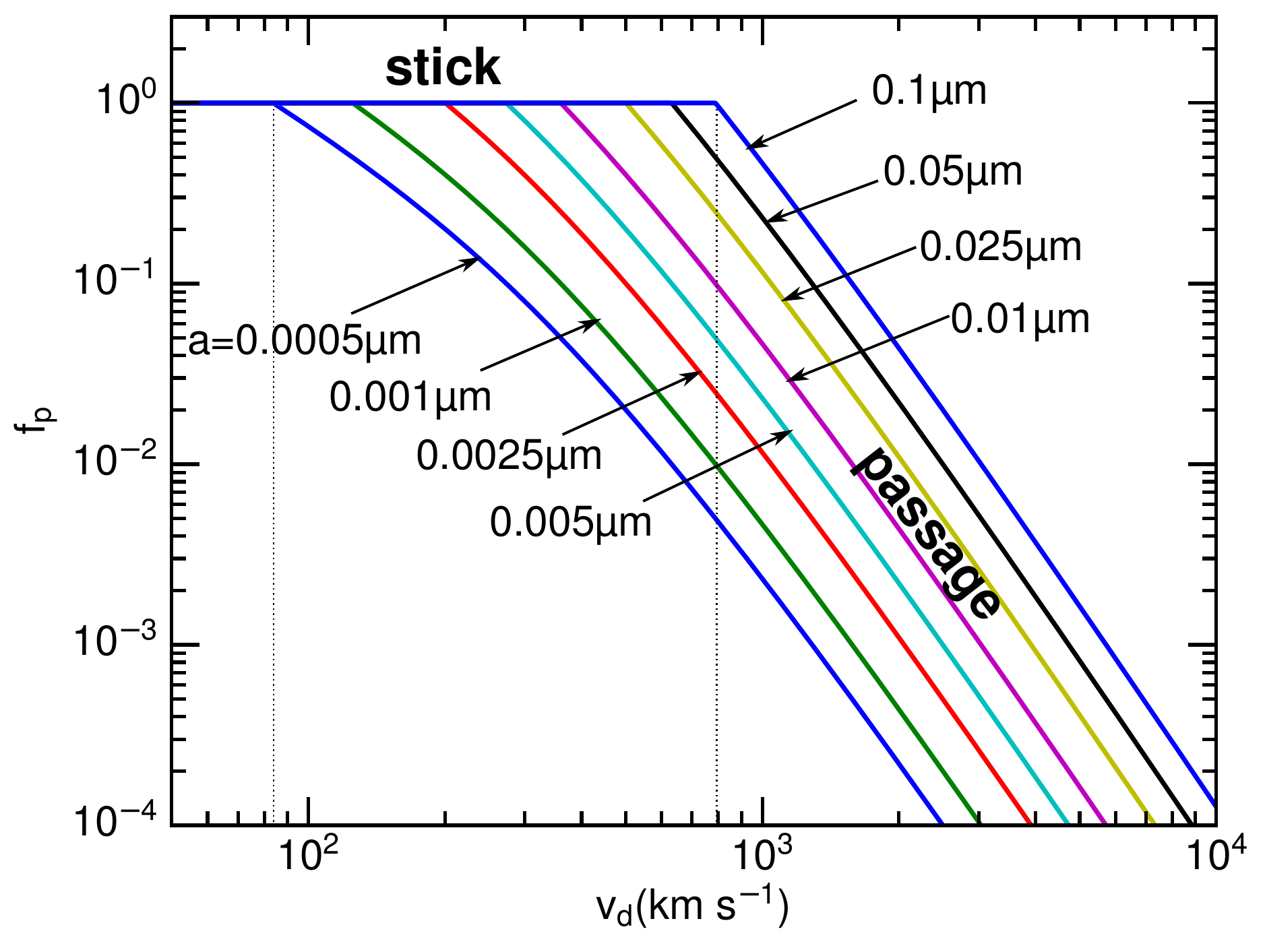}
\caption{The fraction of proton momentum transferred to the grain, $f_{p}$, as a function of grain velocity for the different grain sizes $a$ from $0.0005\mum$ to $0.1\mum$. Graphite material is considered. Stick regime $f_p=1$ and passage regime ($f_{p}<1$) are separated by a vertical dotted line.}
\label{fig:fp}
\end{figure}

To obtain the disruption time by METD, we first calculate the fraction of particle momentum transferred to the grain upon collisions for the different velocity using Equation (\ref{eq:fE}). The obtained results are shown in Figure \ref{fig:fp}. For low velocities ($v_{d}<100\km\s^{-1}$), $f_{p}=1$, which correspond to the sticking that impinging particles collide and stick to the grain. For high velocities, $f_{p}$ decreases rapidly with increasing $v_{d}$ due to the passage of incident particles. The transition velocity from the stick to passage regimes increases with grain size.

We calculate the characteristic timescale of METD using Equations (\ref{eq:tdisr}) and (\ref{eq:tdisr_high}) for the different grain sizes. The drift velocity $v_{d}$ is varied to cover both the low-energy regime (stick) and high-energy (passage) regime. The tensile strength is also varied to reflect different grain structures. We also calculate the timescales of rotational damping by gas collisions and electric dipole emission, gas drag, and sputtering using Equations (\ref{eq:tauH}), (\ref{eq:taued}), (\ref{eq:tdrag}), and (\ref{eq:tau_sp}) and compare the obtained results with the METD time.

\begin{figure*}
\includegraphics[width=0.5\textwidth]{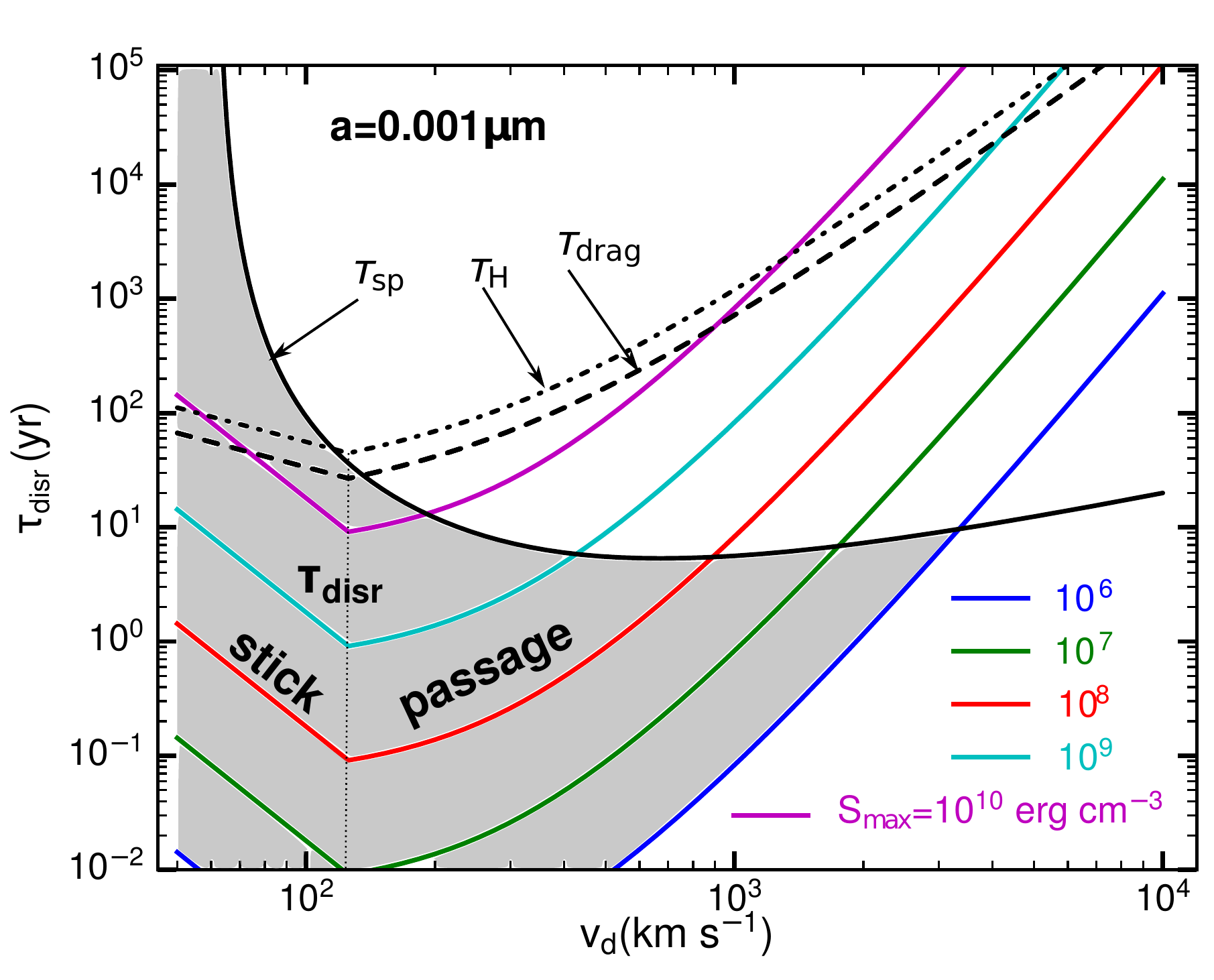}
\includegraphics[width=0.5\textwidth]{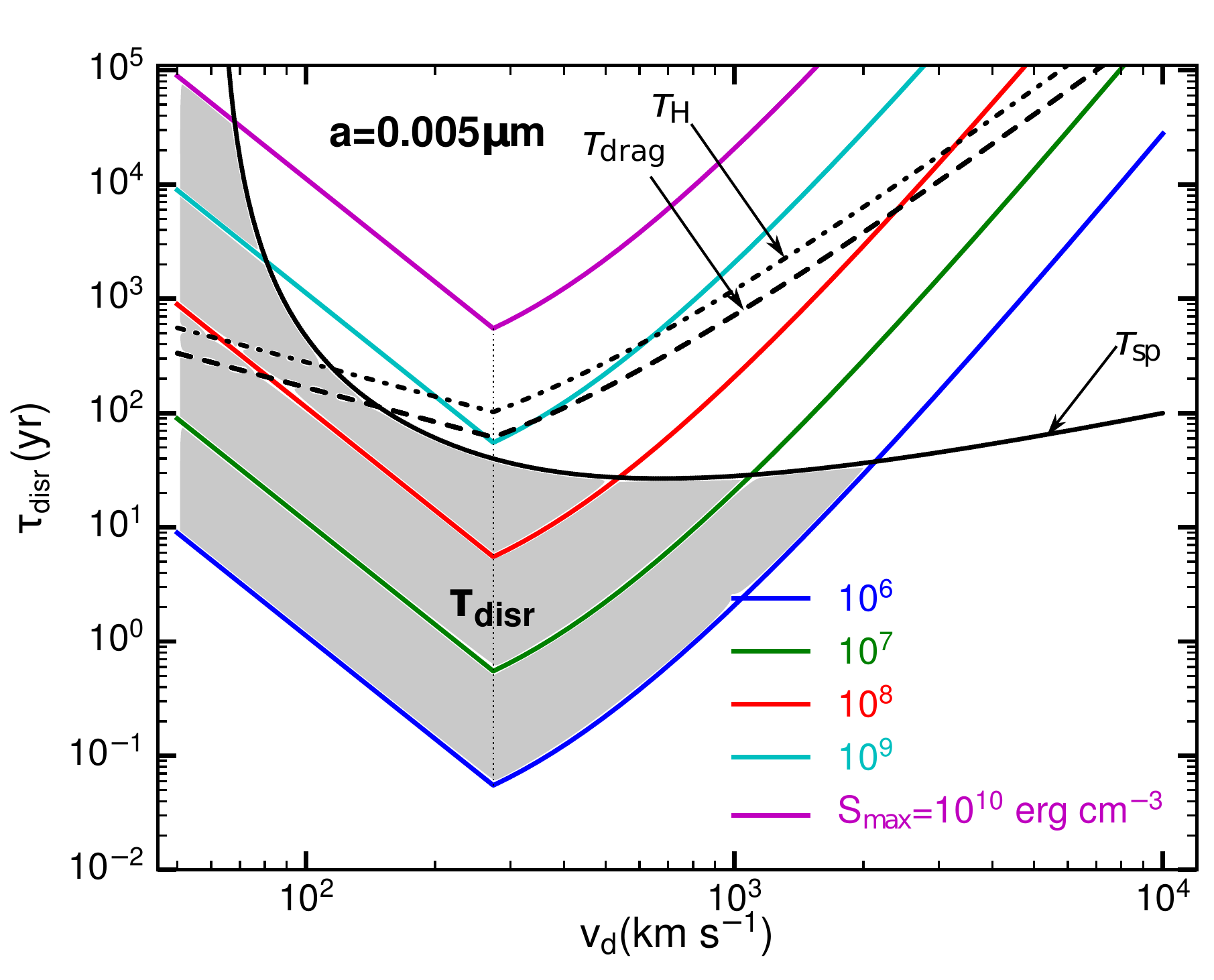}
\includegraphics[width=0.5\textwidth]{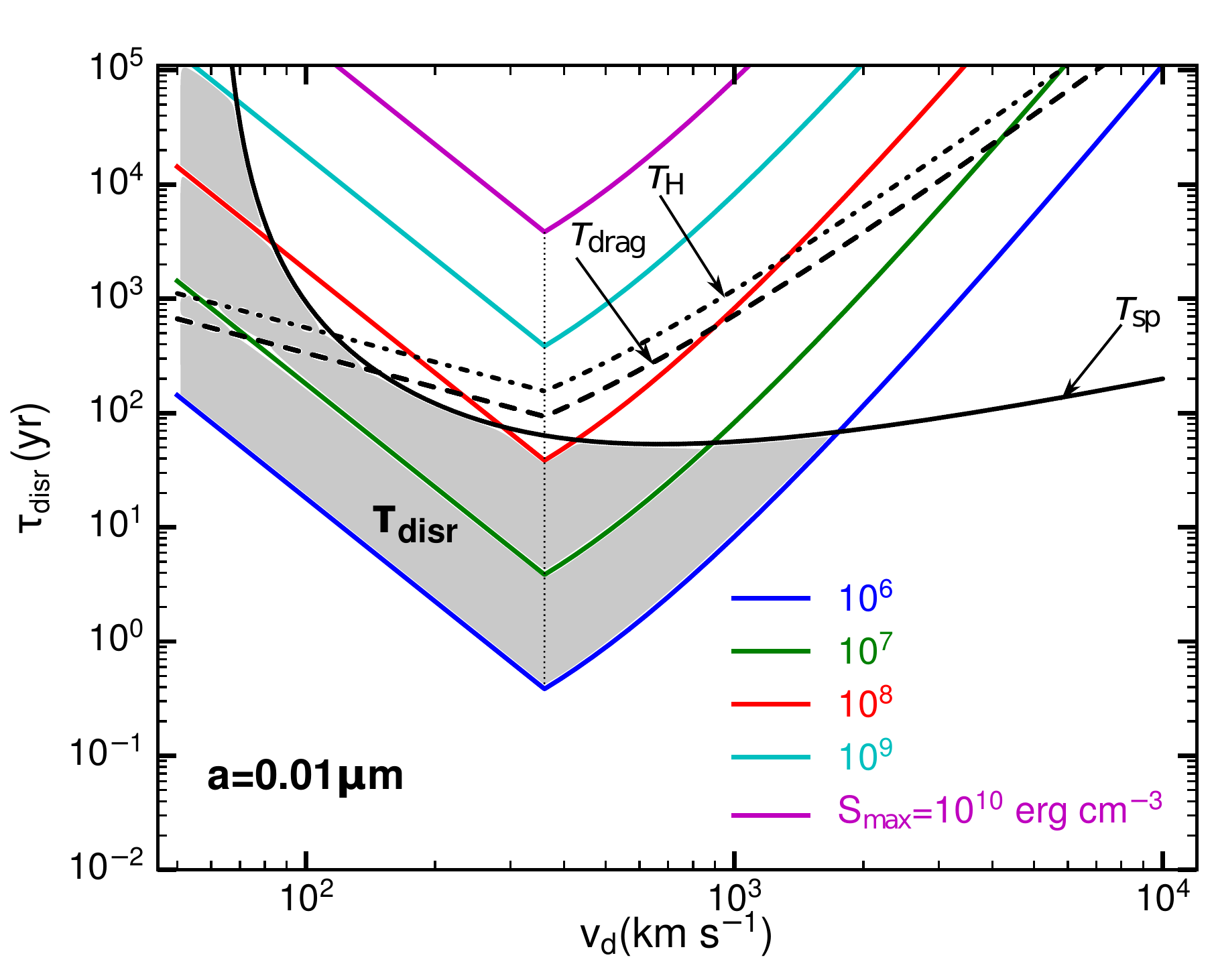}
\includegraphics[width=0.5\textwidth]{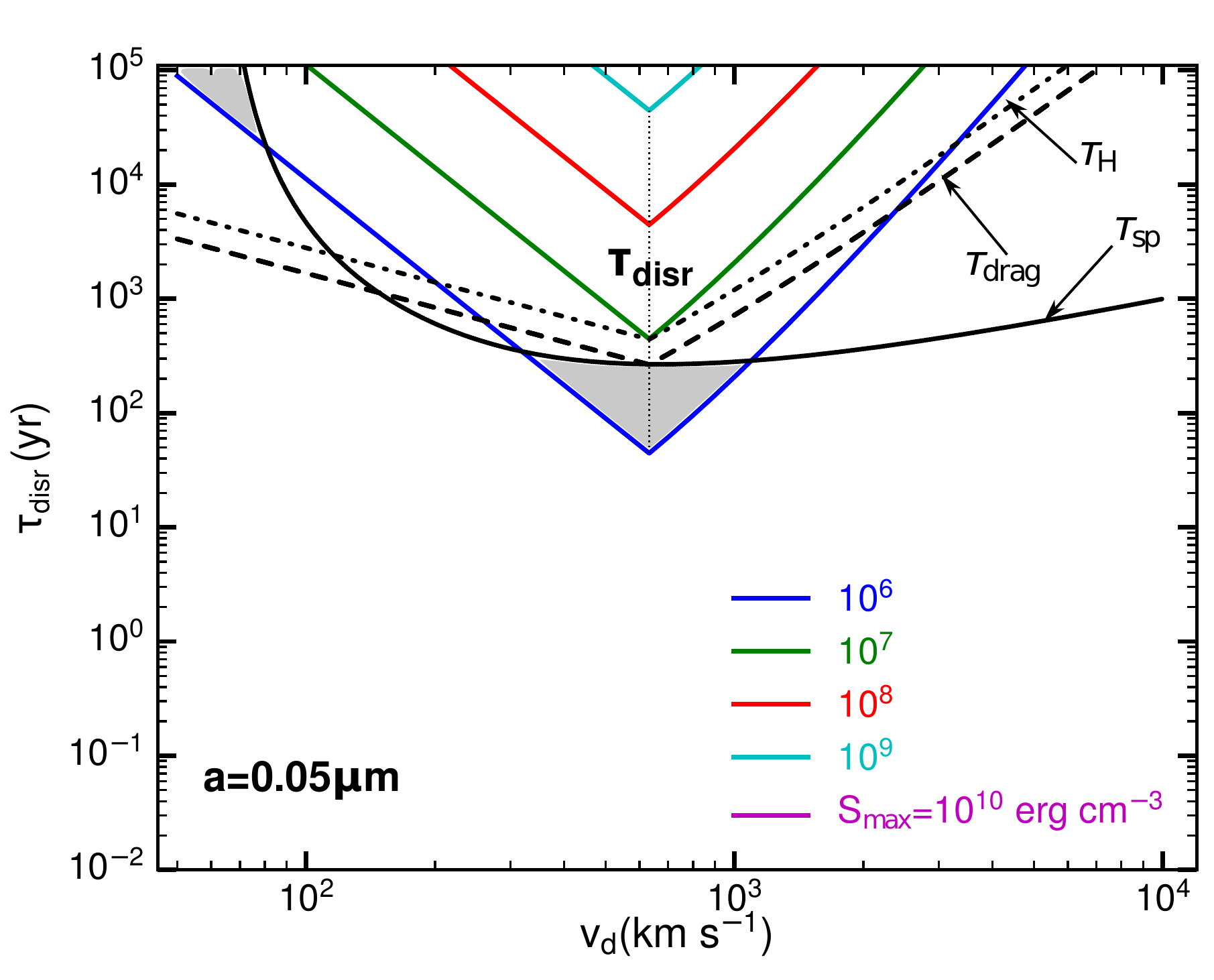}
\caption{The METD timescale vs. the grain velocity for the different grain sizes and tensile strength $S_{\max}$. The timescales of nonthermal sputtering, gas drag, rotational damping, and electric dipole damping are overplotted for comparison. Shaded regions correspond to $\tau_{\rm disr}\le \tau_{\rm sp}$. The vertical line mark the transition from stick to passage regimes.}
\label{fig:time_vd}
\end{figure*}

Figure \ref{fig:time_vd} shows the METD timescale compared to the various timescales, assuming the different grain sizes and tensile strength. The shaded regions mark the parameter space where METD is faster than nonthermal sputtering, characterized by $\tau_{\rm disr}\le \tau_{\rm sp}$. The METD time rapidly decreases with grain velocity $v_{d}$ during the low-energy regime where impinging particles stick to the grain and transfer their entire momentum to the grain ($f_{p}=1$). When $v_{d}$ becomes sufficiently large, impinging particles just pass through the grain, and $\tau_{\rm disr}$ reverses the trend and increases with increasing $v_{d}$ due to the reduction of the ion momentum transfer to the grain (i.e., $f_{p}<1$).

For small grains of $a=0.001\mum$, METD is faster than sputtering for velocities of $v_{d}< 180\km\s^{-1}$ for ideal material of $S_{\max}=10^{10}\erg\cm^{-3}$. For weak materials of $S_{\max}\sim 10^{7}\erg\cm^{-3}$ (e.g., fluffy grains), METD is much faster than sputtering for $v_{d}<1500\km\s^{-1}$. For large grains of $a=0.05\mum$, rotational disruption is only effective for grains with low tensile strength of $S_{\max}\sim 10^{6}\erg\cm^{-3}$ at $v_{d}\sim 300-1000\km\s^{-1}$.

\begin{figure*}
\includegraphics[width=0.5\textwidth]{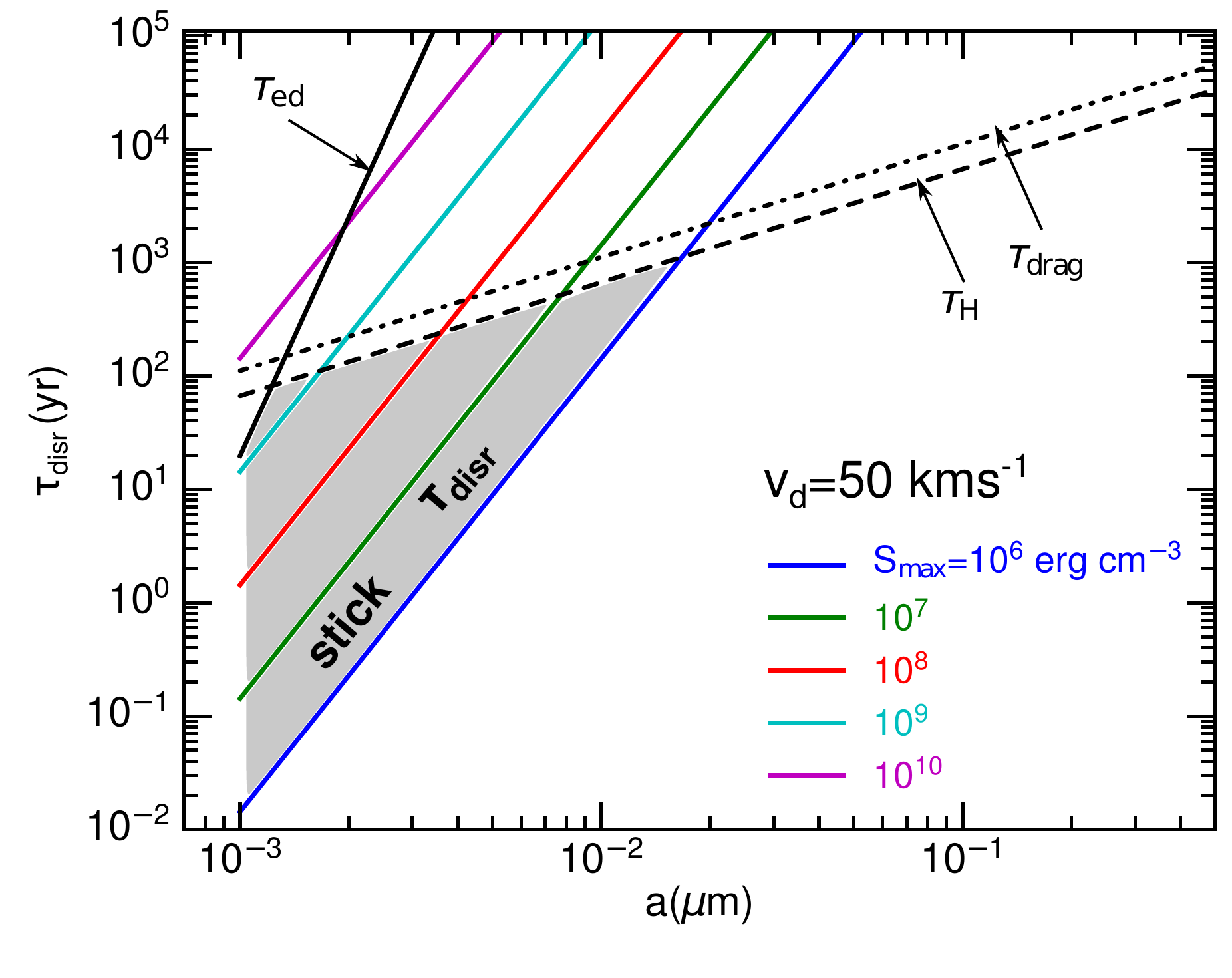}
\includegraphics[width=0.5\textwidth]{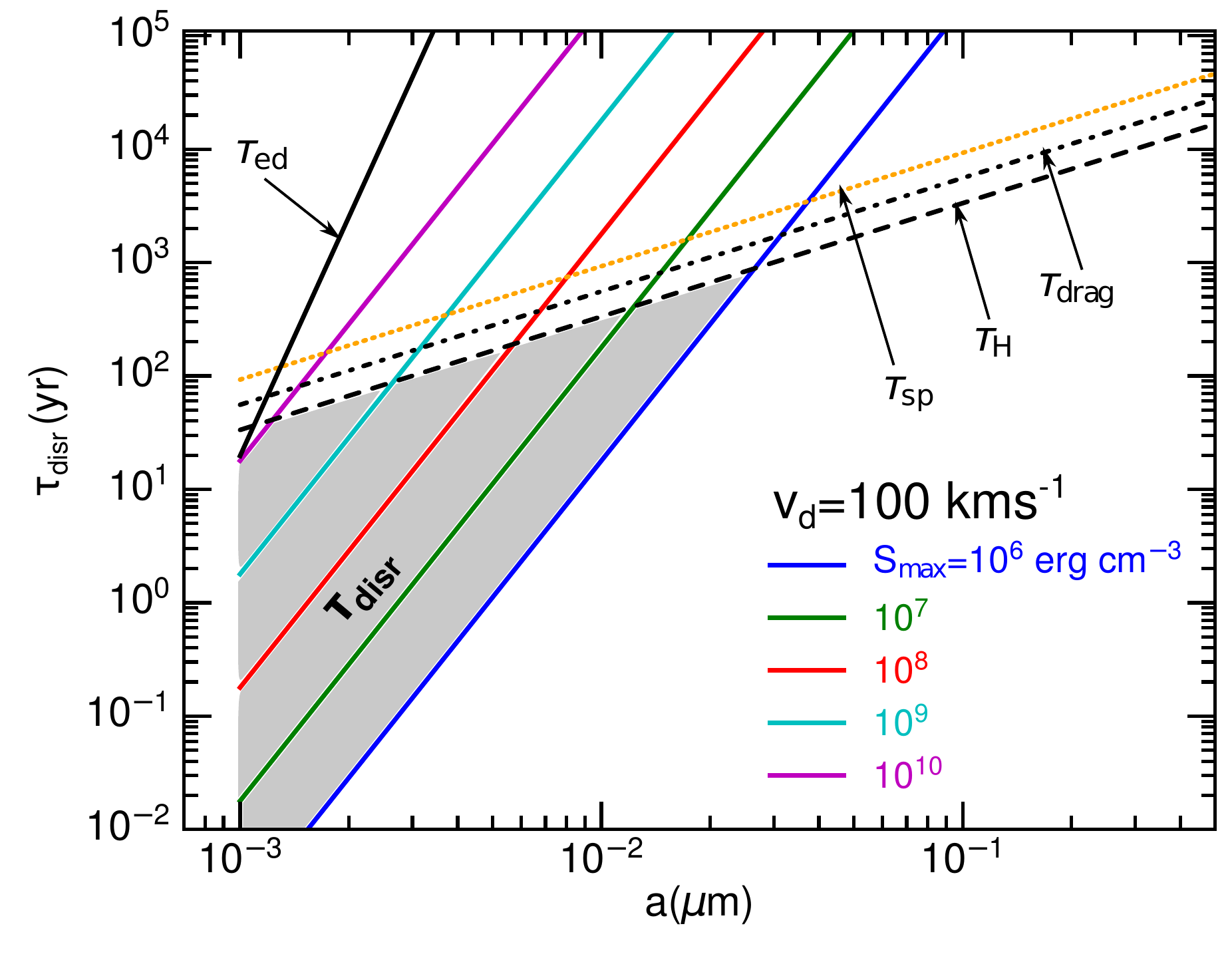}
\includegraphics[width=0.5\textwidth]{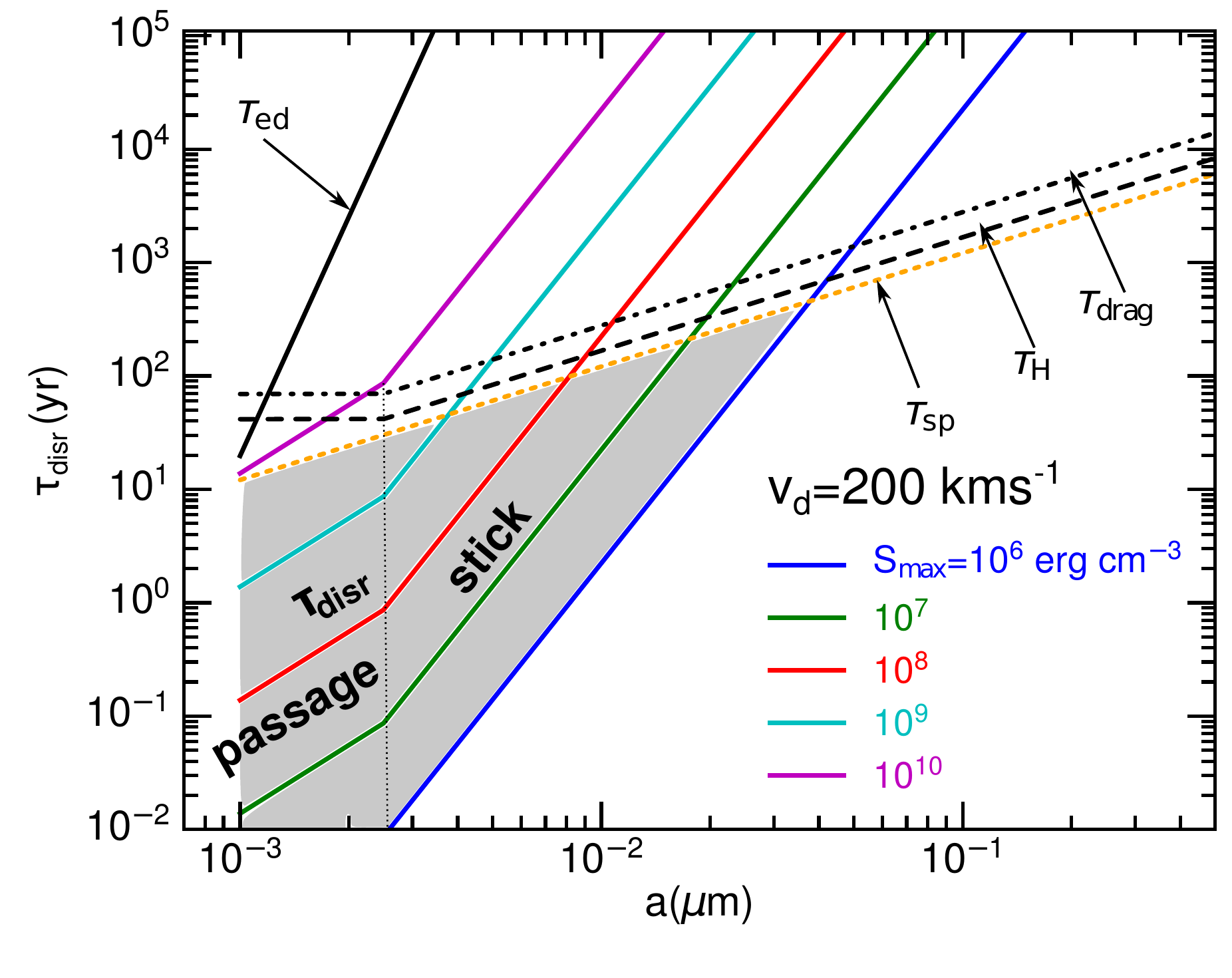}
\includegraphics[width=0.5\textwidth]{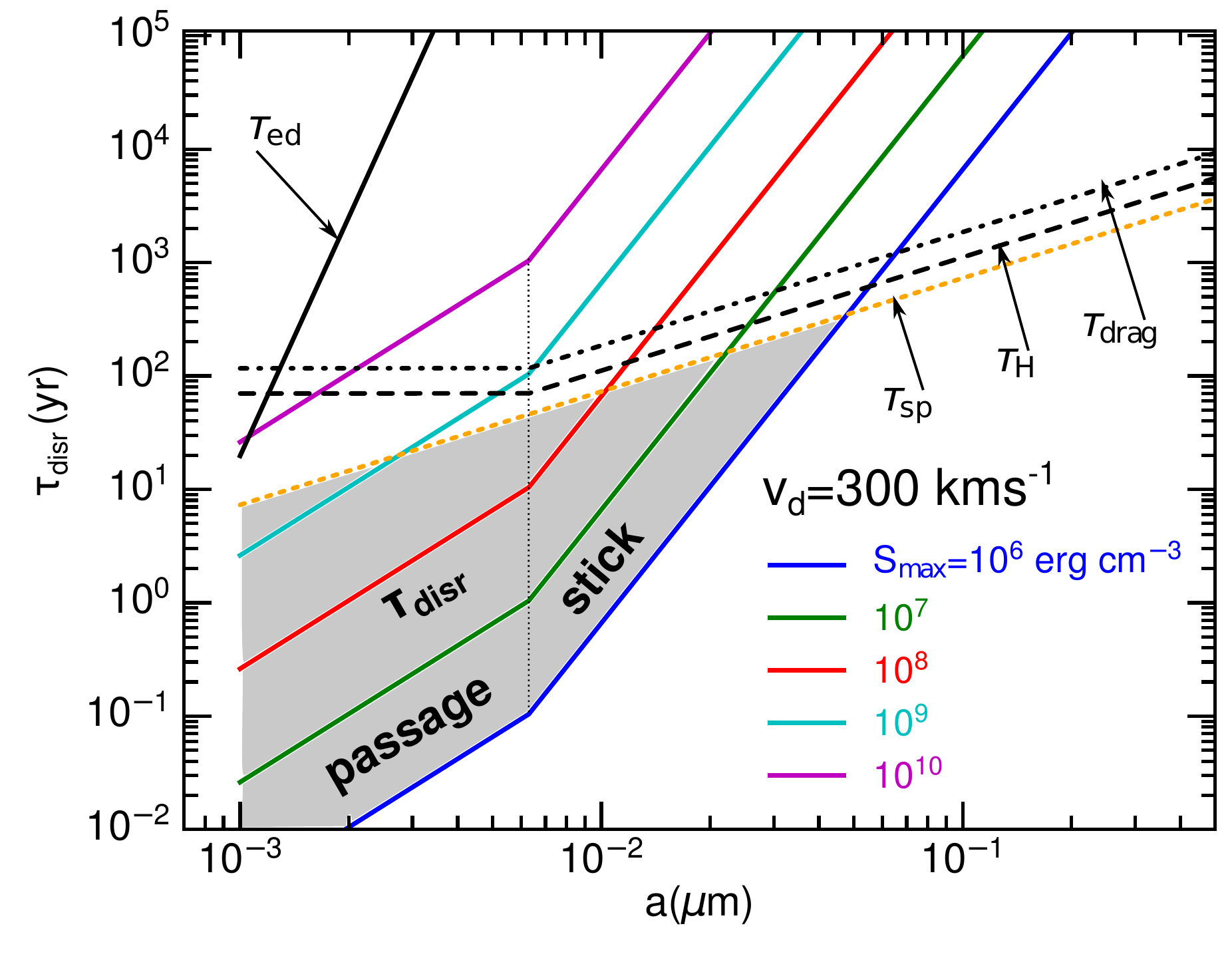}
\caption{Characteristic timescale of METD vs. grain size for the different grain velocities and tensile strength. Timescales of gas drag, rotational damping by gas collisions and electric dipole emission, and nonthermal sputtering are shown for comparison. Shaded areas represent the space where METD is faster than other processes. The vertical line mark the transition from the passage to stick regimes.}
\label{fig:time_size}
\end{figure*}

\begin{figure*}
\includegraphics[width=0.5\textwidth]{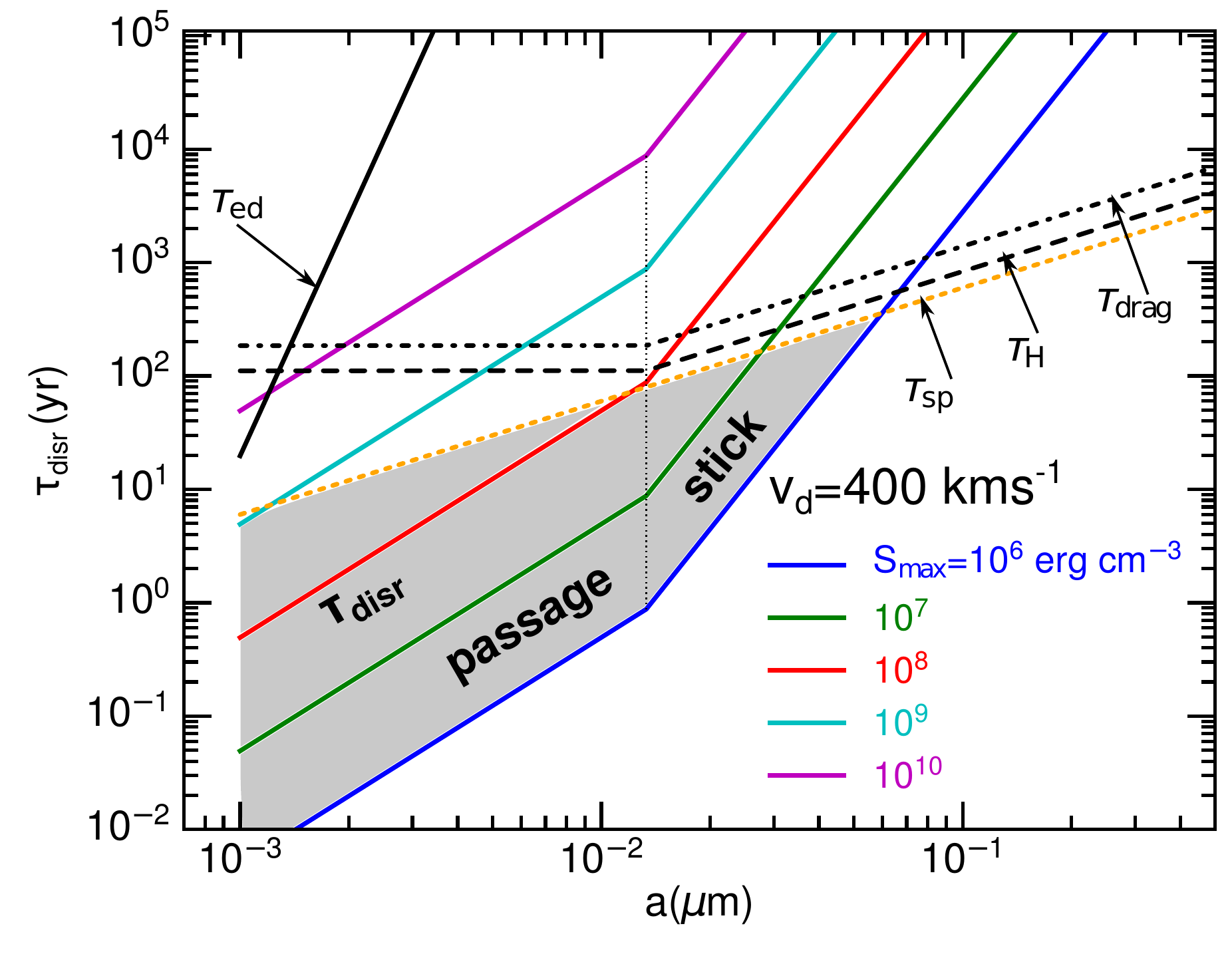}
\includegraphics[width=0.5\textwidth]{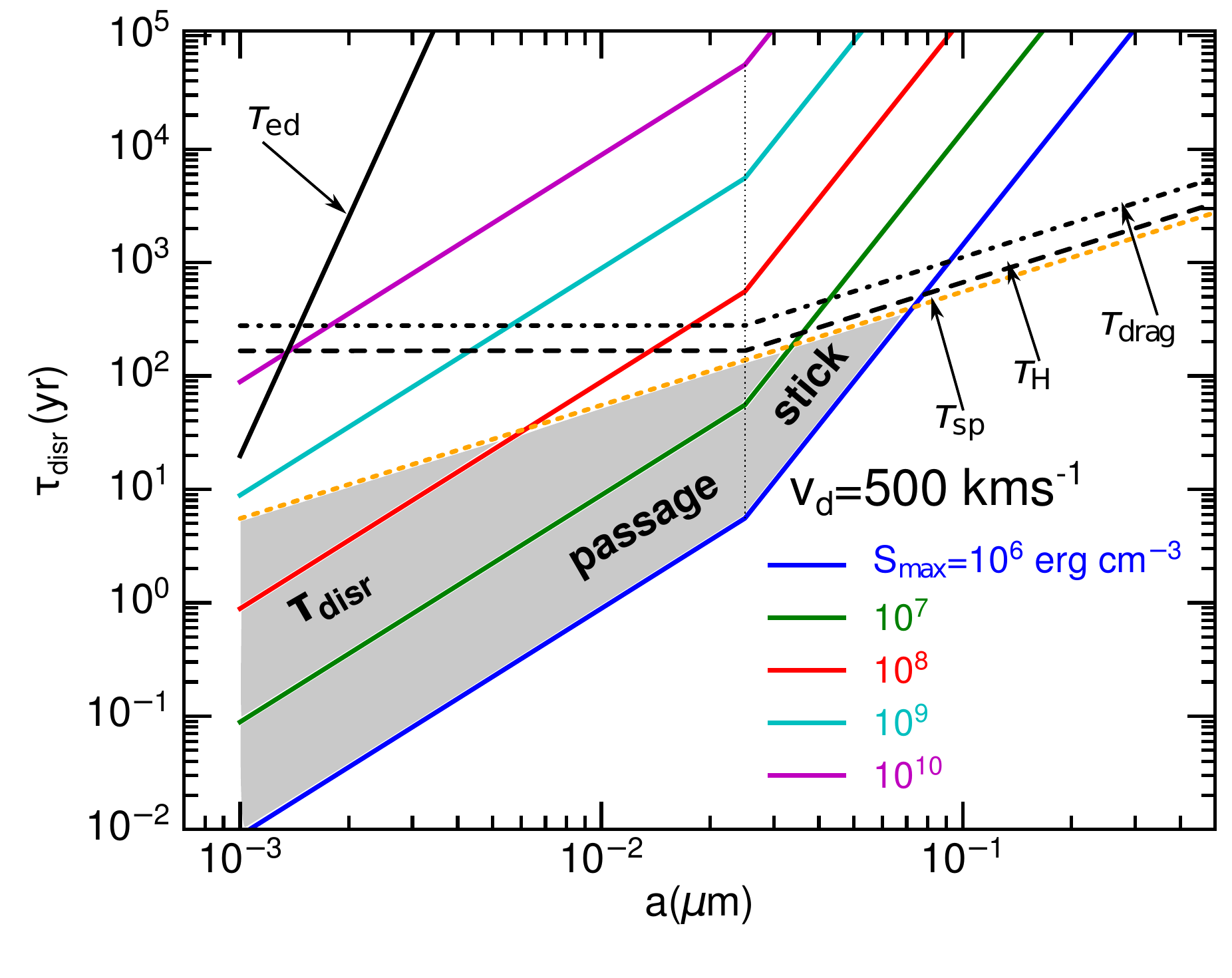}
\includegraphics[width=0.5\textwidth]{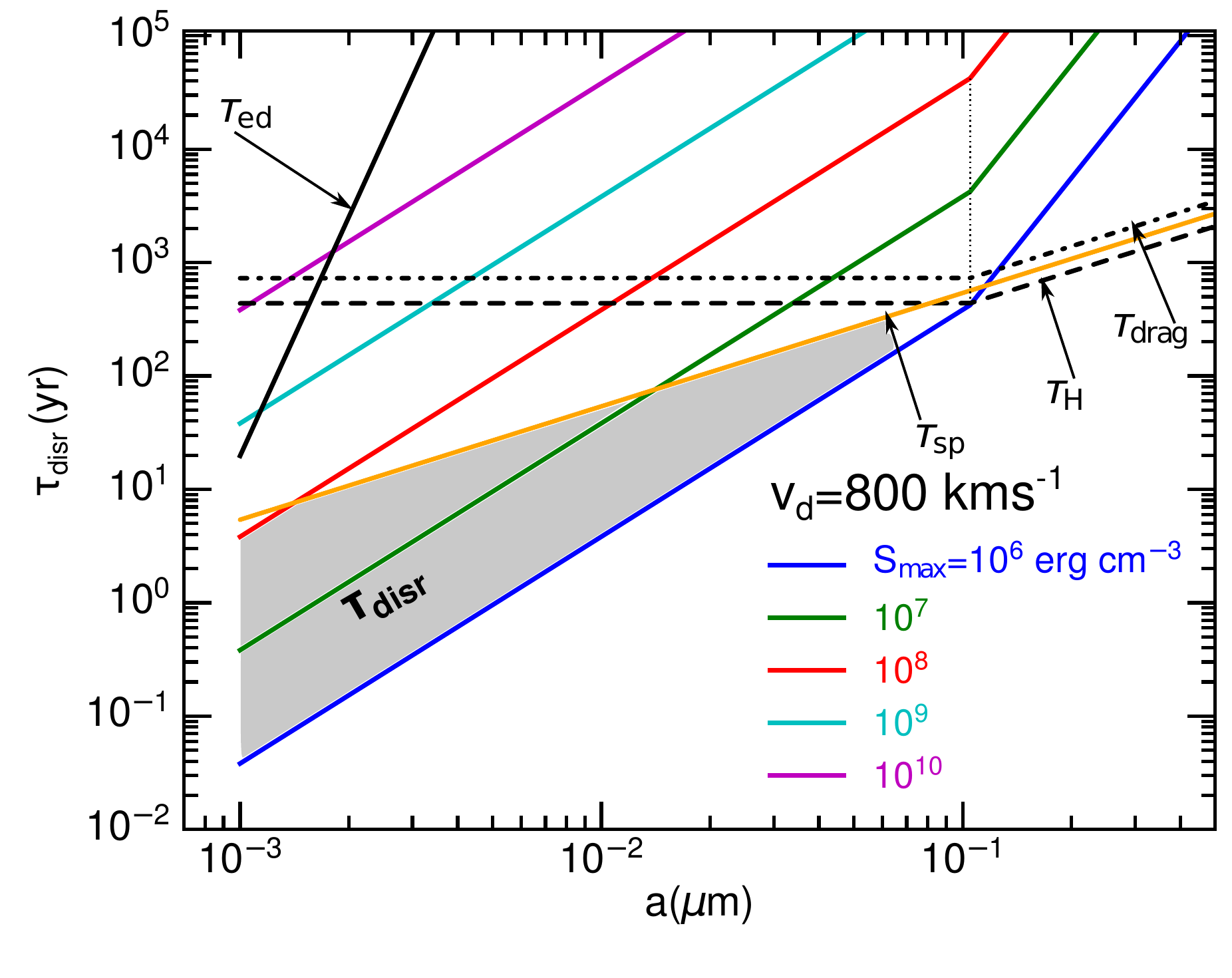}
\includegraphics[width=0.5\textwidth]{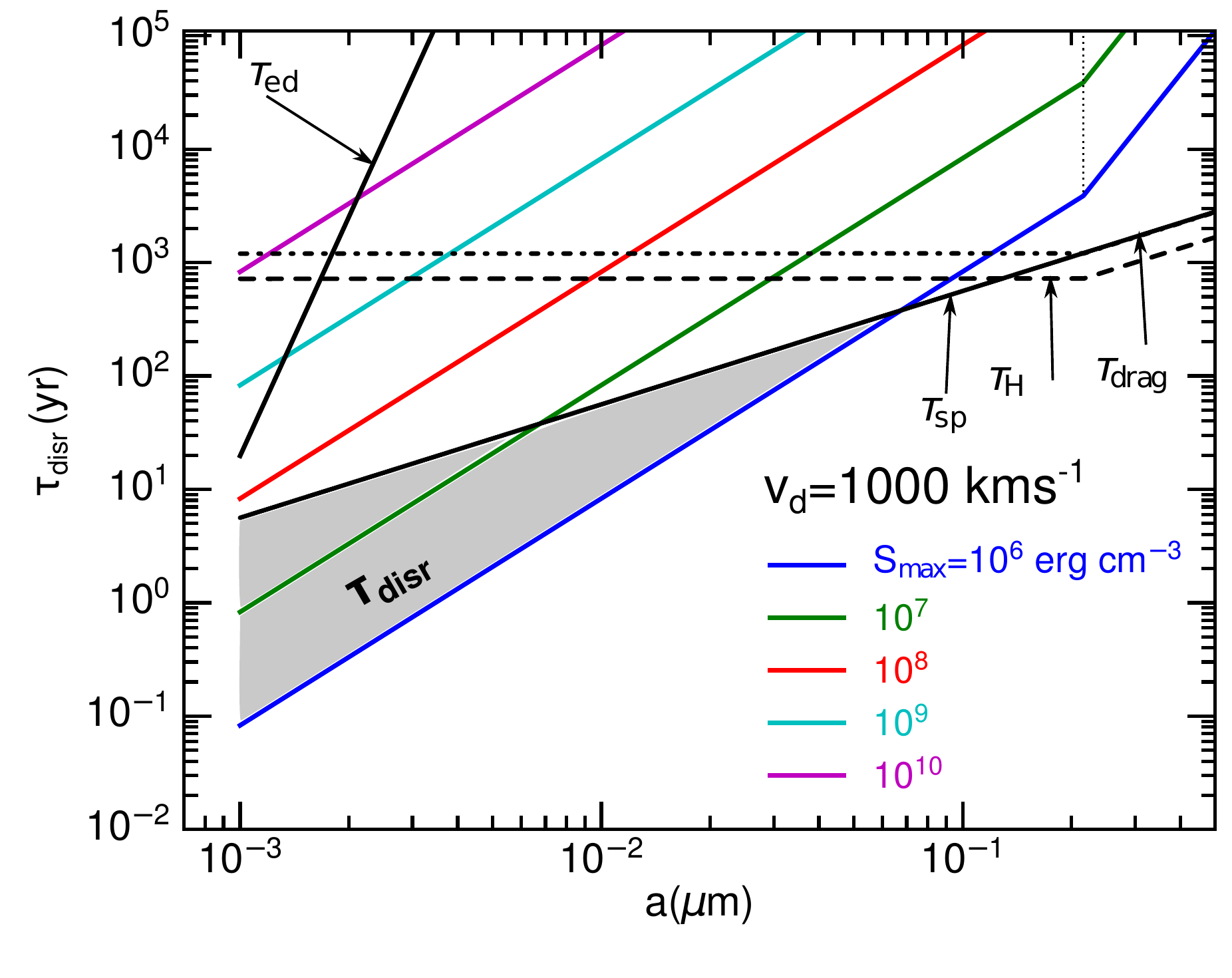}
\caption{Same as Figure \ref{fig:time_vd} but for velocities $v_{d}\ge 400\km\s^{-1}$. METD becomes less efficient due to the degree of ion momentum transfer to the grain but still important for small grains of non-ideal structures.}
\label{fig:time_size_highE}
\end{figure*}

Figures \ref{fig:time_size} shows the METD timescale as a function of the grain size for different tensile strength and $v_{d}=50-300\km\s^{-1}$. METD is faster than gas drag as well as sputtering for small grains and high velocities. At low velocities $v_{d}\le 100\km\s^{-1}$, impinging particles collide and stick to the grain surface, and $\tau_{\rm disr}$ decreases rapidly with grain size (upper panels). At higher velocities, incident particles can pass through the grain if their sizes are small (see lower panels), resulting in the change in the slope of $\tau_{\rm disr}$ vs. $a$.

Figure \ref{fig:time_size_highE} shows the similar result for higher velocities of $v_{d}=400-1000 \km\s^{-1}$. The range of grain sizes where METD is efficient is narrower for higher $v_{d}$. This arises from the decrease of ion momentum transfer to the grain at very high $v_{d}$ (see Figure \ref{fig:fp}).

From Equations (\ref{eq:tdisr}) and (\ref{eq:tau_sp}) one obtains the ratio of METD time to nonthermal sputtering time:
\bea
\frac{\tau_{\rm disr}}{\tau_{\rm sp}}\simeq 6.4S_{\max,9}\left(\frac{\bar{A}_{\rm sp}}{12}\right)\left(\frac{Y_{\rm sp}}{0.1}\right)\left(\frac{a_{-6}^{3}}{v_{2}^{2}}\right)f_{p}^{-2},\label{eq:disr_sput}
\ena
where $f_{p}=1$ for low velocities but decreases rapidly with increasing $v_{d}$ at very high velocities (see Eq. \ref{eq:fE}).

Equation (\ref{eq:disr_sput}) reveals that the METD mechanism is faster than nonthermal sputtering for small grains (e.g., nanoparticles). Indeed, from Equation (\ref{eq:delta_omega}) one can see that a single collision can spin up the grain to $\delta \omega\sim 10^{7}a_{-6}^{-4}v_{3}\rad\s^{-1}$. Thus, to spin-up the $0.01\mum$ grain to the disruption limit, $\omega_{\rm cri}\sim 10^{9}\rad\s^{-1}S_{\max,9}^{1/2}$, it only requires $N_{\rm disr}\sim (\omega_{\rm cri}/\delta \omega)^{2}\sim 10^{4}$ random collisions. However, nonthermal sputtering of yield $Y_{\rm sp}\sim 0.1$ requires $N_{at}/Y_{\rm sp}\sim 10^{6}$ collisions to completely destroy the grain where $N_{at}$ is the total number of atoms in the dust grain.

\subsection{Grain disruption sizes}
\begin{figure*}
\includegraphics[width=0.5\textwidth]{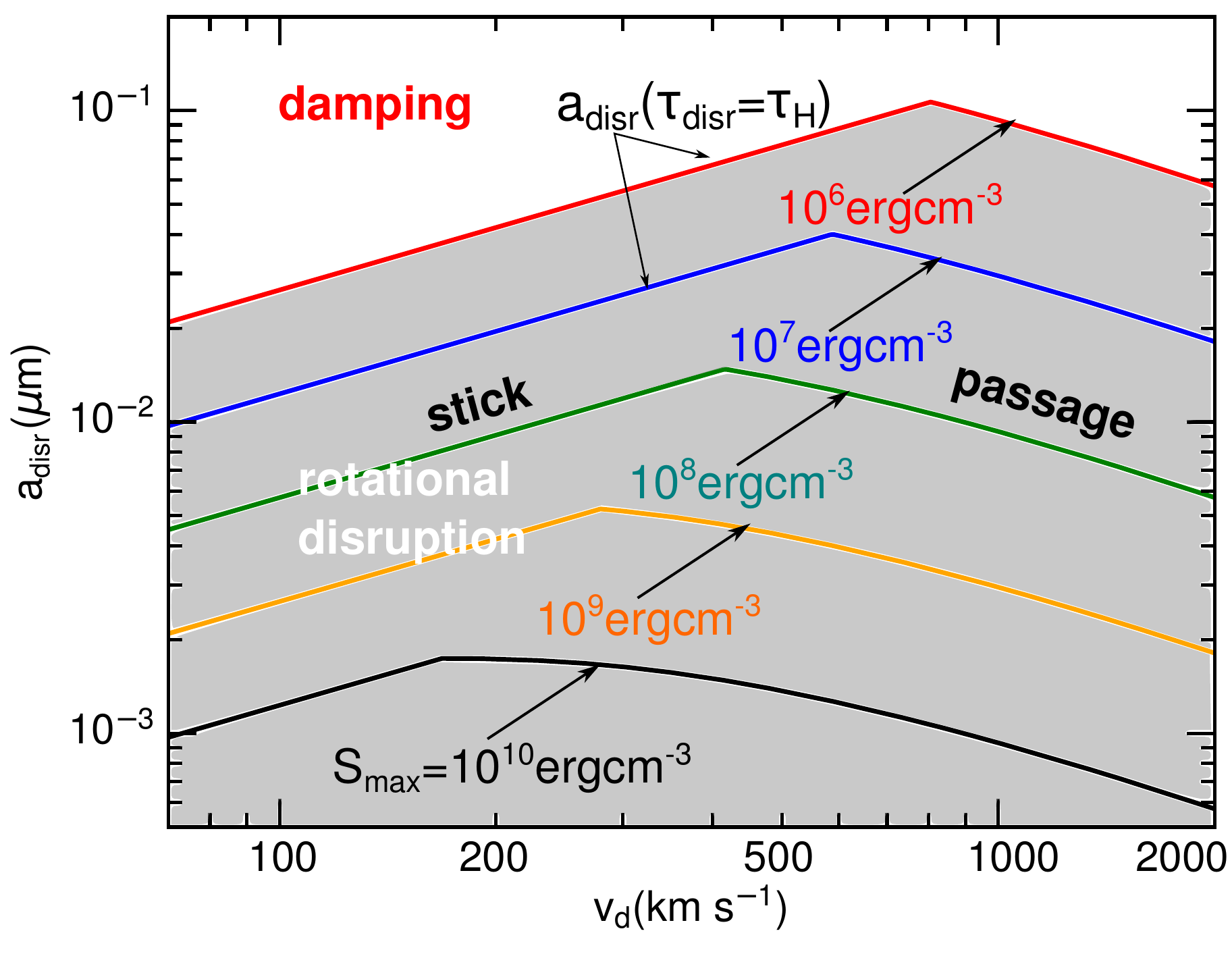}
\includegraphics[width=0.5\textwidth]{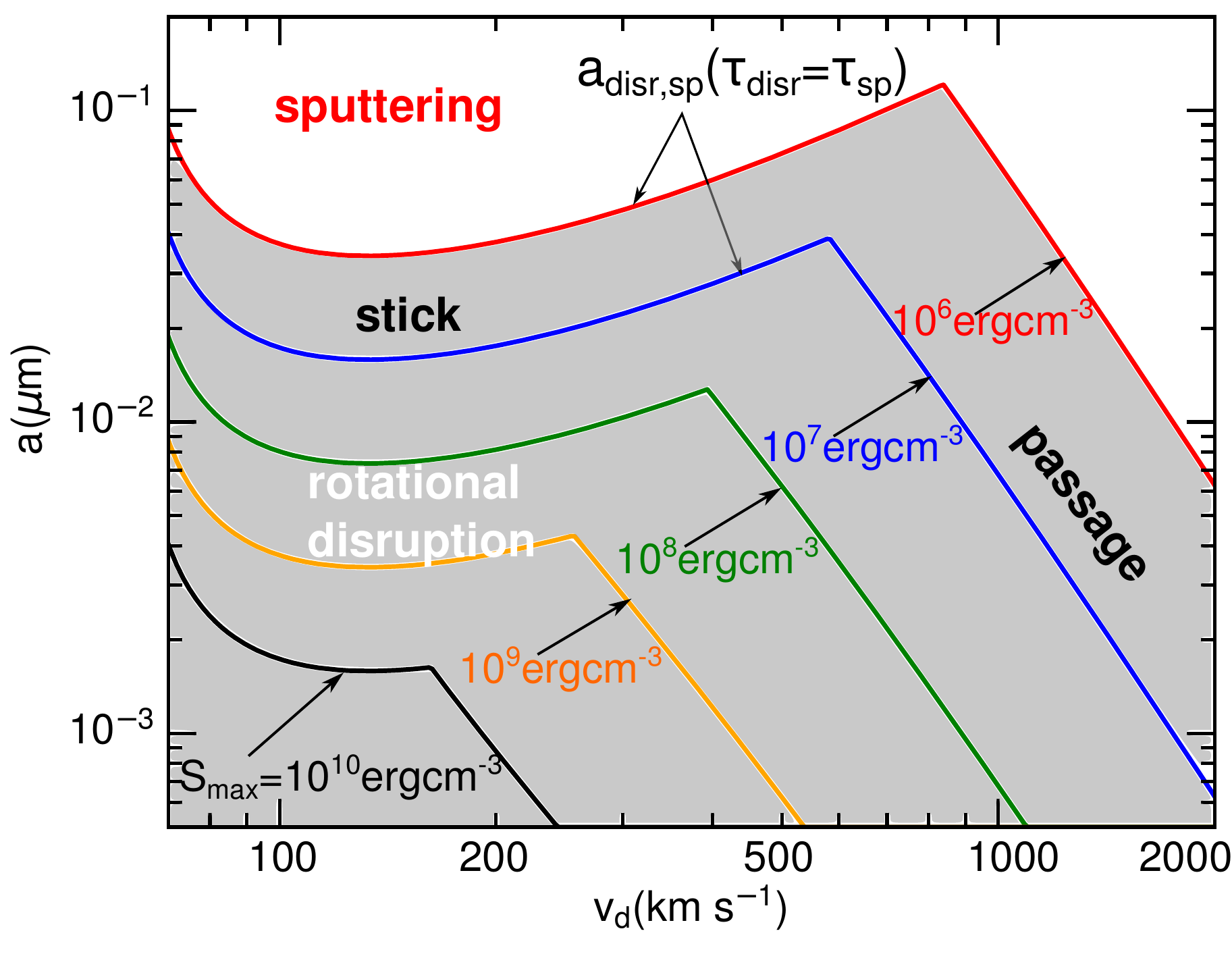}
\caption{Grain disruption size vs. grain velocity assuming the different tensile strength of grain material. The solid lines mark the boundary between rotational disruption and damping ($\tau_{\rm disr}=\tau_{\H}$) and $\tau_{\rm disr}=\tau_{\rm sp}$ (right panel). Shaded areas mark the parameter space where rotational disruption (METD) is faster than rotational damping (left panel) and nonthermal sputtering (right panel).}
\label{fig:adisr}
\end{figure*}

To obtain grain disruption size $a_{\rm disr}$ for arbitrary velocities $v_{d}$, we first calculate $\tau_{\rm disr}$ for a range of grain sizes and compare with rotational damping time $\tau_{\H}$. As shown in Figures \ref{fig:time_size} and \ref{fig:time_size_highE}, METD becomes faster than nonthermal sputtering for grains smaller than the value at the intersection (i.e., $\tau_{\rm disr}=\tau_{\rm sp}$), which is denoted by $a_{\rm disr,sp}$. 

Figure \ref{fig:adisr} (left panel) shows the values of $a_{\rm disr}$ as a function of grain velocity for various tensile strength. The disruption size $a_{\rm disr}$ increases with increasing $v_{d}$ and then decreases due to the decrease of ion momentum transfer to the grain ($f_{p}<1$).

Figure \ref{fig:adisr} (right panel) shows the variation of $a_{\rm disr,sp}$ with $v_{d}$. Shaded areas mark the parameter space ($v_{d},a$) in which METD is faster than nonthermal sputtering. For weak grains (e.g., of fluffy structure) with $S_{\max}\sim 10^{7}\erg\cm^{-3}$, grains of $a\sim 0.02\mum$ can be disrupted for $v_{d}<600\km\s^{-1}$. For very strong grains of ideal structures with $S_{\max}\sim 10^{9}\erg\cm^{-3}$, very small grains of $a\sim 0.004\mum$ can be rotational disrupted for $v_{d}<250\km\s^{-1}$. Nonthermal sputtering dominates the destruction of large grains (i.e., $a>0.1\mum$) or at high velocities of $v_{d}>1000\km\s^{-1}$.

\section{Discussion}\label{sec:discuss}
\subsection{The importance of grain internal structures for dust destruction by gas-grain collisions}
Internal structures (compact vs. porous) of dust grains are essential for dust absorption and emission (\citealt{Guillet:2017hg}). The grain internal structure determines the mechanical strength of grains as measured by tensile strength. Indeed, grain internal structures are found to play an important role for grain coagulation \citep{1997ApJ...480..647D}. However, this property is previously ignored in the destruction process by gas-grain collisions (thermal and nonthermal processes).

Very recently, with the discovery of RAdiative Torque Disruption (RATD) mechanism \citep{Hoang:2019da}, the grain tensile strength is crucially important in determining the upper cutoff of the grain size distribution (\citealt{2019ApJ...876...13H}). Our present study reveals that the tensile strength also plays a critical role in dust destruction by gas-grain collisions. Small grains ($a<0.01\mum$) of fluffy structures with low tensile strength ($S_{\max}<10^{9}\erg\cm^{-3}$) can be destroyed more efficiently than nonthermal sputtering when moving at high velocities through the gas. While nonthermal sputtering does depend on the internal structure of grains (e.g., only on the binding energy of target atoms), rotational disruption critically depends on the tensile strength $S_{\rm max}$. Very small grains ($a\lesssim 1 nm$) can be destroyed by METD even when they have ideal structures.

\subsection{Destruction of grains accelerated by radiation pressure and implications for IGM}
Dust grains can be accelerated to hypersonic velocities by radiation pressure from intense radiation sources such as massive OB stars, supernovae (SNe), and Active Galactic Nuclei (see, e.g., \citealt{2017ApJ...847...77H}).

The maximum gas column density that a small grain of size $a<a_{\rm disr}$ can traverse before being disrupted by rotational disruption is estimated as
\bea
N_{\rm max,disr}&=& n_{\rm H}v_{d} \tau_{\rm disr}\nonumber\\
&\simeq& 1.5\times 10^{18} a_{-6}^{4}S_{\rm max,9}v_{2}^{-2}f_{p}^{-2}~\cm^{-2}.\label{eq:Nmax}
\ena

Similarly, one can evaluate the maximum distance determined by sputtering as
\bea
N_{\rm max,sp}=n_{\rm H}v_{d} \tau_{\rm sp}\simeq 7.6\times 10^{18} a_{-6}Y_{\rm sp,-1}~\cm^{-2},
\ena
where $Y_{sp,-1}=Y_{\rm sp}/(0.1)$.

The grain can only traverse a distance of $D_{\rm max,disr}=N_{max,disr}/n_{\rm H} \sim 10^{4}(50/n_{1})S_{\rm max,9}a_{-6}^{4}v_{2}^{-2}$ AU before being disrupted by METD. However, if small grains can move at extreme speeds of $v_{d}>1000\km\s^{-1}$, the effect of METD becomes less efficient due to the decrease of ion momentum transfer to the grain (i..e, $f_{p}\ll 1)$. 

Observations show the presence of dust in the circumgalactic medium (CGM) and intergalactic medium (IGM). The underlying mechanism to the existence of dust in the CGM and IGM is due to radiation pressure that expels grains at high speeds. The enrichment of metals and dust in the IGM is believed to arise from two main routes, including galactic winds driven by supernovae and the injection of galactic grains moving at high velocities of $v_{d}>100\km\s^{-1}$ accelerated by radiation pressure (see e.g., \citealt{2005MNRAS.358..379B}). \cite{1991ApJ...381..137F} find that radiation pressure from starlight can accelerate grains to $v_{d}\sim 100-600\km\s^{-1}$ from the Galaxy to galactic halo over a timescale of Myr. At such high velocities, our results show that small grains of sizes $a<0.01\mum$ are efficiently destroyed by METD (see Figure \ref{fig:adisr}, right panel) if they have non-ideal structures (i.e., composite or porous structures).


\subsection{Mechanical torque disruption of dust grains in fast shocks}
Shocks are ubiquitous in the interstellar medium, which includes slow shocks of velocities $v_{\rm sh}<50\km\s^{-1}$ driven by the outflow of young stars, stellar winds, and jets, and fast shocks of velocities $v_{\rm sh}>100\km\s^{-1}$ driven by supernova blast waves.

Grain shattering dominates for slow shocks (\citealt{1979ApJ...231..438D}; see \citealt{1995Ap&SS.233..111D}), but for fast shocks, sputtering is dominant in destroying refractory grains of small sizes (\citealt{1994ApJ...431..321T}).

A new understanding of dust destruction in shocks is presented in \cite{2019ApJ...877...36H} and \cite{2019ApJ...886...44T}. For steady-state shocks, \cite{2019ApJ...877...36H}, for the first time, found that nanoparticles of size $a<2$ nm can be disrupted into molecules by mechanical torques due to stochastic gas bombardment (i.e., METD mechanism). For the same grain size, sputtering is found to have a lower destruction rate. In this paper, we found that rotational disruption is dominant over sputtering for fast shocks if grains are small or made of weak materials. 

Fast shocks are very common in core-collapse supernova due to the interaction of ejecta with the surrounding environment (see e.g., \citealt{2006ApJ...648..435N}). Newly formed dust grains in the supernova ejecta are subject to reverse shocks and move at velocity of $2/(\gamma+3)v_{\rm sh}$ with adiabatic index $\gamma=5/3$ relative to the gas (see e.g., \citealt{1996ApJ...457..244D}; \citealt{2007MNRAS.378..973B}). Traditionally, thermal and non-thermal sputtering is established to be a main destruction mechanism of dust, which controls the dust evolution in the early universe \citep{2006ApJ...648..435N}. Based on our present calculations, METD appears to dominate over sputtering for fast shocks (i.e., $v_{\rm sh}>100\km\s^{-1}$), especially for small grains with fluffy structures that have low tensile strength. Therefore, newly formed dust in  supernova remnants (SNRs) is expected to have deficiency of small and very small grains of size $a<a_{\rm disr}$ due to rotational disruption, which is in agreement with observations (see \citealt{Micelotta:2018gl} and reference therein). 

Observational data by \cite{2006ApJ...652L..33W} and \cite{2019arXiv190706213Z} show that for the fraction of dust destructed in fast shocks of core-collapse supernova is higher than theoretical predictions based on sputtering. The authors appeal to the porous structure of dust grains to enhance the sputtering rate. However, this observational puzzle can be explained by rotational disruption.
 
\subsection{Rotational disruption cascade of small grains by gas bombardment}
What is the subsequent evolution of fragments resulting from METD due to supersonic motion of grains through the gas? In the frame work of METD, resulting fragments can rotate at higher rotational rates than the original grain due to their smaller masses. As a result, they would be disrupted rapidly into smaller fragments. This results in collisional cascade of grains into smaller and smaller fragments. Finally, nonthermal sputtering acts to destroy such tiny fragments into individual atoms/molecules. The rotational disruption cascade is expected to begin with the RATD mechanism \citep{Hoang:2019da} because small grains can survive against RATD due to their weaker radiative torques.

Finally, we note that, in this paper, spherical grains are considered and we disregarded the effect of regular mechanical torques acting on grains of irregular shapes (\citealt{2007ApJ...669L..77L}; \citealt{2016MNRAS.457.1958D}; \citealt{2018ApJ...852..129H}). The latter is expected to be more efficient.

\subsection{Mechanical torque disruption in hot gas}
In the post-shock regions or ionized plasma, gas can be heated to very high temperatures of $T_{\gas}>10^{6}\K$. In such a hot gas, thermal collisions with protons can spin-up dust grains to extremely fast rotation with $\omega_{\rm rot}=\omega_{T}$, resulting in the rotational disruption (\citealt{1979ApJ...231...77D}). Using the disruption criteria $\omega_{\rm rot}=\omega_{\rm disr}$, one obtains the disruption size for a given $T_{\gas}$ as follows:
\bea
a_{\rm disr}=\left(\frac{45kT_{\gas}}{32\pi S_{\max}} \right)^{1/3}\simeq 0.02 S_{\max,9}^{-1/3}\left(\frac{T_{\gas}}{10^{8}\K}\right)^{1/3}\mum.~~~~
\ena

The critical gas temperature required for rotational disruption in a hot plasma is given by
\bea
T_{\gas}\gtrsim \left(\frac{32\pi a^{3}}{45k}\right)S_{\max}
\simeq1.6\times10^{7}a_{-6}^{3}S_{\rm max,9}\K,
\ena
which implies that the rotational disruption is important for $T_{\gas}\sim 10^{6}\K$ if grains are as small as $0.01\mum$ and made of weak materials ($S_{\max}<10^{9}\erg\cm^{-3}$).

\begin{figure}
\includegraphics[width=0.5\textwidth]{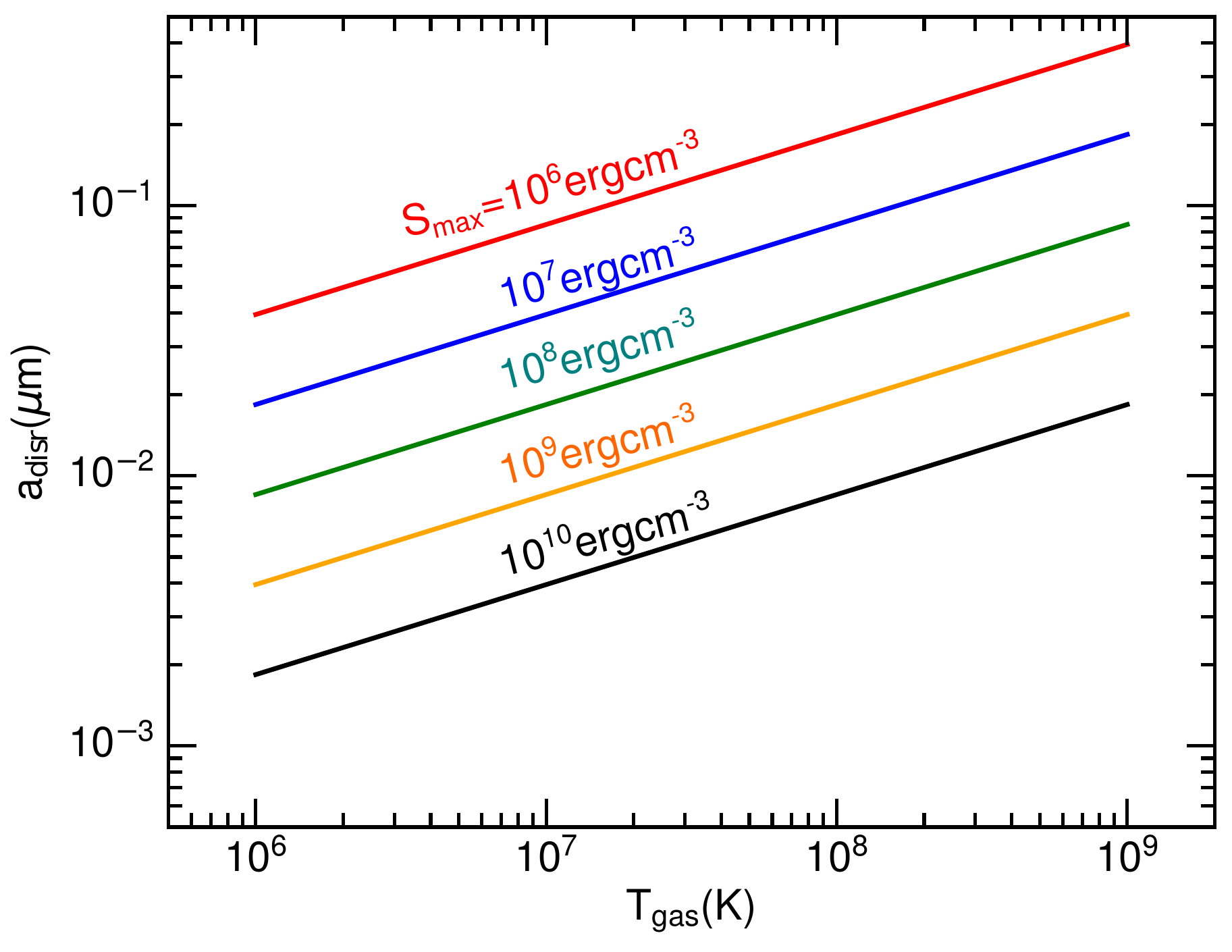}
\caption{Grain disruption size as a function of the gas temperature in hot plasma assuming the different tensile strength.}
\label{fig:adisr_hotgas}
\end{figure}

Figure \ref{fig:adisr_hotgas} shows the disruption size as a function of the gas temperature. Small grains are destroyed by rotational disruption so that dust grains in hot plasma are only larger than $a_{\rm disr}$. Large grains with weak structures can be destroyed by rotational disruption for temperatures $T_{\rm gas}\sim 10^{9}\K$. So, we predict that dust grains in very hot gas likely have a compact structure of high tensile strength.

\section{Summary}\label{sec:summary}
We study rotational disruption of dust grains by stochastic mechanical torques due to gas-grain collisions at high velocities. Our results are summarized as follows:

\begin{enumerate}

\item We find that very small and small grains can be disrupted by centrifugal stress within rapidly spinning grains due to stochastic gas-grain collisions when the relative grain velocity is sufficiently large. The exact velocity threshold for disruption is determined by the tensile strength of dust grains.
  
\item We compare the timescale of METD with sputtering timescale and find that rotational disruption is much more efficient than sputtering for grain velocities $v_{d}<500 \km\s^{-1}$. The ratio of rotational disruption to sputtering time is $\tau_{\rm disr}/\tau_{\rm sp}\sim 0.7S_{\max,9}Y_{sp,-1}a_{-6}^{3}(\bar{A}_{\rm sp}/12)(300\km\s^{-1}/v_{d})^{-2}$. The internal structure of grains which determines the tensile strength is found to play a critical role in grain destruction by gas-grain collisions.

\item At very high velocities (e.g., $v_{d}>500\km\s^{-1}$), we find that the rate of METD for small grains is reduced by a factor $f_{p}^{2}$ because incident particles pass through the grain and transfer only a fraction $f_{p}$ of their momentum to the grain.

\item We discuss the implications of our study for the origin of intergalactic dust and find that small grains are likely disrupted before injecting into the IGM. Large grains of $a\sim 0.05-0.1\mum$ of weak material may be disrupted, but grains of strong material such as compact grains can survive and reach the IGM.

\item Our results demonstrate that METD may be a dominant destruction mechanism of small grains in fast shocks of $v_{\rm sh}>100\km\s^{-1}$, instead of non-thermal sputtering as previously thought. Very small grains can be disrupted at lower velocities. This enhanced dust destruction rate appears to be consistent with dust destruction measured from fast shocks of core-collapse supernovae.

\item We also find that METD can also be more efficient than thermal sputtering for small grains of weak material in hot gas, assuming that grains rotate at thermal angular velocity.

\end{enumerate}
 
\acknowledgements
We are grateful to the referee, Dr. Joseph Nuth, for a constructive review that improves our manuscript. We thank A. Lazarian for his warm encouragements and Sergio M. Gonzalez for helpful comments. This work was supported by the National Research Foundation of Korea (NRF) grants funded by the Korea government (MSIT) through the Basic Science Research Program (2017R1D1A1B03035359) and Mid-career Research Program (2019R1A2C1087045).

\bibliographystyle{/Users/thiemhoang/Dropbox/Papers2/apj}
\bibliography{ms.bbl}

\begin{thebibliography}{43}
\expandafter\ifx\csname natexlab\endcsname\relax\def\natexlab#1{#1}\fi

\bibitem[{Aguirre {et~al.}(2001{\natexlab{a}})Aguirre, Hernquist, Katz,
  Gardner, \& Weinberg}]{2001ApJ...556L..11A}
Aguirre, A., Hernquist, L., Katz, N., Gardner, J., \& Weinberg, D.
  2001{\natexlab{a}}, \apj, 556, L11

\bibitem[{Aguirre {et~al.}(2001{\natexlab{b}})Aguirre, Hernquist, Schaye, Katz,
  Weinberg, \& Gardner}]{2001ApJ...561..521A}
Aguirre, A., Hernquist, L., Schaye, J., {et~al.} 2001{\natexlab{b}}, \apj, 561,
  521

\bibitem[{Bianchi \& Ferrara(2005)}]{2005MNRAS.358..379B}
Bianchi, S., \& Ferrara, A. 2005, \mnras, 358, 379

\bibitem[{Bianchi \& Schneider(2007)}]{2007MNRAS.378..973B}
Bianchi, S., \& Schneider, R. 2007, \mnras, 378, 973

\bibitem[{Bohdansky(1984)}]{1984NIMPB...2..587B}
Bohdansky, J. 1984, \nimpb, 2, 587

\bibitem[{Das \& Weingartner(2016)}]{2016MNRAS.457.1958D}
Das, I., \& Weingartner, J.~C. 2016, \mnras, 457, 1958

\bibitem[{Dominik \& Tielens(1997)}]{1997ApJ...480..647D}
Dominik, C., \& Tielens, A. G. G.~M. 1997, \apj, 480, 647

\bibitem[{Draine(1995)}]{1995Ap&SS.233..111D}
Draine, B.~T. 1995, Astrophysics and Space Science, 233, 111

\bibitem[{Draine \& Lazarian(1998)}]{1998ApJ...508..157D}
Draine, B.~T., \& Lazarian, A. 1998, \apj, 508, 157

\bibitem[{Draine \& Salpeter(1979{\natexlab{a}})}]{1979ApJ...231..438D}
Draine, B.~T., \& Salpeter, E.~E. 1979{\natexlab{a}}, \apj, 231, 438

\bibitem[{Draine \& Salpeter(1979{\natexlab{b}})}]{1979ApJ...231...77D}
Draine, B.~T., \& Salpeter, E.~E. 1979{\natexlab{b}}, \apj, 231, 77

\bibitem[{Dwek {et~al.}(1996)Dwek, Foster, \& Vancura}]{1996ApJ...457..244D}
Dwek, E., Foster, S.~M., \& Vancura, O. 1996, \apj, 457, 244

\bibitem[{Ellison {et~al.}(1997)Ellison, Drury, \& Meyer}]{1997ApJ...487..197E}
Ellison, D.~C., Drury, L.~O., \& Meyer, J.-P. 1997, \apj, 487, 197

\bibitem[{Epstein(1980)}]{1980MNRAS.193..723E}
Epstein, R.~I. 1980, \mnras, 193, 723

\bibitem[{Ferrara {et~al.}(1991)Ferrara, Ferrini, Franco, \&
  Barsella}]{1991ApJ...381..137F}
Ferrara, A., Ferrini, F., Franco, J., \& Barsella, B. 1991, Astrophysical
  Journal, 381, 137

\bibitem[{Gold(1952)}]{1952MNRAS.112..215G}
Gold, T. 1952, \mnras, 112, 215

\bibitem[{Goldreich \& Scoville(1976)}]{1976ApJ...205..144G}
Goldreich, P., \& Scoville, N. 1976, \apj, 205, 144

\bibitem[{Guillet {et~al.}(2017)Guillet, Fanciullo, Verstraete, Boulanger,
  Jones, Miville-Desch{\^e}nes, Ysard, Levrier, \& Alves}]{Guillet:2017hg}
Guillet, V., Fanciullo, L., Verstraete, L., {et~al.} 2017, \aa

\bibitem[{Hoang(2017)}]{2017ApJ...847...77H}
Hoang, T. 2017, \apj, 847, 77

\bibitem[{Hoang(2019)}]{2019ApJ...876...13H}
Hoang, T. 2019, \apj, 876, 13

\bibitem[{Hoang {et~al.}(2018)Hoang, Cho, \& Lazarian}]{2018ApJ...852..129H}
Hoang, T., Cho, J., \& Lazarian, A. 2018, \apj, 852, 129

\bibitem[{Hoang {et~al.}(2010)Hoang, Draine, \& Lazarian}]{Hoang:2010jy}
Hoang, T., Draine, B.~T., \& Lazarian, A. 2010, \apj, 715, 1462

\bibitem[{Hoang {et~al.}(2012)Hoang, Lazarian, \& Schlickeiser}]{Hoang:2012cx}
Hoang, T., Lazarian, A., \& Schlickeiser, R. 2012, \apj, 747, 54

\bibitem[{Hoang {et~al.}(2015)Hoang, Lazarian, \&
  Schlickeiser}]{2015ApJ...806..255H}
Hoang, T., Lazarian, A., \& Schlickeiser, R. 2015, \apj, 806, 255

\bibitem[{Hoang \& Tram(2019)}]{2019ApJ...877...36H}
Hoang, T., \& Tram, L.~N. 2019, \apj, 877, 36

\bibitem[{Hoang {et~al.}(2019)Hoang, Tram, Lee, \& Ahn}]{Hoang:2019da}
Hoang, T., Tram, L.~N., Lee, H., \& Ahn, S.-H. 2019, Nature Astronomy, 3, 766

\bibitem[{Hoang {et~al.}(2016)Hoang, Vinh, \& Quynh~Lan}]{2016ApJ...824...18H}
Hoang, T., Vinh, N.~A., \& Quynh~Lan, N. 2016, \apj, 824, 18

\bibitem[{Ishibashi \& Fabian(2015)}]{Ishibashi:2015bu}
Ishibashi, W., \& Fabian, A.~C. 2015, \mnras, 451, 93

\bibitem[{Jones {et~al.}(1994)Jones, Tielens, Hollenbach, \&
  McKee}]{1994ApJ...433..797J}
Jones, A.~P., Tielens, A. G. G.~M., Hollenbach, D.~J., \& McKee, C.~F. 1994,
  \apj, 433, 797

\bibitem[{Laki{\'c}evi{\'c} {et~al.}(2015)Laki{\'c}evi{\'c}, van Loon, Meixner,
  Gordon, Bot, Roman-Duval, Babler, Bolatto, Engelbracht, Filipovi{\'c}, Hony,
  Indebetouw, Misselt, Montiel, Okumura, Panuzzo, Patat, Sauvage, Seale,
  Sonneborn, Temim, Uro{\v s}evi{\'c}, \& Zanardo}]{2015ApJ...799...50L}
Laki{\'c}evi{\'c}, M., van Loon, J.~T., Meixner, M., {et~al.} 2015, \apj, 799,
  50

\bibitem[{Lazarian \& Hoang(2007)}]{2007ApJ...669L..77L}
Lazarian, A., \& Hoang, T. 2007, \apj, 669, L77

\bibitem[{Micelotta {et~al.}(2018)Micelotta, Matsuura, \&
  Sarangi}]{Micelotta:2018gl}
Micelotta, E.~R., Matsuura, M., \& Sarangi, A. 2018, Space Sci Rev, 1

\bibitem[{Netzer \& Elitzur(1993)}]{1993ApJ...410..701N}
Netzer, N., \& Elitzur, M. 1993, \apj, 410, 701

\bibitem[{Nozawa {et~al.}(2006)Nozawa, Kozasa, \& Habe}]{2006ApJ...648..435N}
Nozawa, T., Kozasa, T., \& Habe, A. 2006, \apj, 648, 435

\bibitem[{Purcell \& Spitzer(1971)}]{1971ApJ...167...31P}
Purcell, E.~M., \& Spitzer, L.~J. 1971, \apj, 167, 31

\bibitem[{Sankrit {et~al.}(2010)Sankrit, Williams, Borkowski, Gaetz, Raymond,
  Blair, Ghavamian, Long, \& Reynolds}]{2010ApJ...712.1092S}
Sankrit, R., Williams, B.~J., Borkowski, K.~J., {et~al.} 2010, \apj, 712, 1092

\bibitem[{{Sigmund}(1981)}]{1981spb1.book....9S}
{Sigmund}, P. 1981, in Sputtering by Particle Bombardment I, ed. R.~{Behrisch}
  (New York: Springer), 9

\bibitem[{Silvia {et~al.}(2010)Silvia, Smith, \& Shull}]{2010ApJ...715.1575S}
Silvia, D.~W., Smith, B.~D., \& Shull, J.~M. 2010, \apj, 715, 1575

\bibitem[{Tielens {et~al.}(1994)Tielens, McKee, Seab, \&
  Hollenbach}]{1994ApJ...431..321T}
Tielens, A. G. G.~M., McKee, C.~F., Seab, C.~G., \& Hollenbach, D.~J. 1994,
  \apj, 431, 321

\bibitem[{Tram \& Hoang(2019)}]{2019ApJ...886...44T}
Tram, L.~N., \& Hoang, T. 2019, \apj, 886, 44

\bibitem[{Williams {et~al.}(2006)Williams, Borkowski, Reynolds, Blair,
  Ghavamian, Hendrick, Long, Points, Raymond, Sankrit, Smith, \&
  Winkler}]{2006ApJ...652L..33W}
Williams, B.~J., Borkowski, K.~J., Reynolds, S.~P., {et~al.} 2006, \apj, 652,
  L33

\bibitem[{Yan {et~al.}(2004)Yan, Lazarian, \& Draine}]{Yan:2004ko}
Yan, H., Lazarian, A., \& Draine, B.~T. 2004, \apj, 616, 895

\bibitem[{Zhu {et~al.}(2019)Zhu, Slane, Raymond, \& Tian}]{2019arXiv190706213Z}
Zhu, H., Slane, P., Raymond, J., \& Tian, W.~W. 2019, arXiv:1907.06213,
  arXiv:1907.06213

\end{thebibliography}

\appendix
\section{Validity of the slab approximation for the momentum transfer}\label{apdx:slab}
In Section \ref{sec:disr}, we assumed a slab approximation to calculate the fraction of the momentum transfer of projectiles to the grain by high-energy collisions. Subsequently, the stochastic torques are calculated as in the case of low-energy collisions.

We now consider whether this assumption is valid. Let us consider an energetic ion bombarding the spherical grain surface at position described by polar angle $\theta$. The ion energy transferred to the grain is in general given by
\bea
\delta E(\theta)= (2a\sin\theta)nS(E),~\label{eq:deltaE1}
\ena
where the pathlength of the projectile in the grain is $2a\sin\theta$. 

The corrected fraction of the momentum transfer to the grain is then equal to
\bea
f_{p,\rm corr}(E,a)\equiv \frac{\delta p}{p}=\frac{\delta E}{2E}=\frac{a\sin\theta}{E}nS(E)=\left(\frac{3\sin\theta}{2}\right)f_{p}(E,a),\label{eq:fE1}
\ena
where $f_{p}(E,a)=(2a/3E)nS(E)$ is the momentum transfer obtained from the slab approximation (see Equation (\ref{eq:fE}).

The increase in the squared angular momentum becomes
\bea
(\delta J)^{2}=(a\cos\theta \delta p)^{2}=(a\cos\theta pf_{p,\rm corr})^{2}=\frac{9a^{2}p^{2}\sin^{2}\theta\cos^{2}\theta}{4}f_{p}(E,a)^{2}=\frac{9a^{2}p^{2}\sin^{2}(2\theta)}{8}f_{p}(E,a)^{2}.
\ena

By averaging the above equation over the grain surface (angle $\theta$ from $0-\pi$), one obtains
\bea
\langle (\delta J)^{2}\rangle = \left(\frac{a^{2}p^{2}}{2}\right)f_{p}(E,a)
\times \left(\frac{9}{8}\right).\label{eq:dJ2_hinew}
\ena

Comparing Equation (\ref{eq:dJ2_hinew}) with (\ref{eq:dJ2_hi}) one can see that the exact calculation is larger than the slab approximation by a small factor of $9/8$.

\end{document}